\documentclass[12pt,aps,floats,showpacs]{revtex4}
\usepackage{graphicx}
\usepackage{epsfig}

\newcommand{\be}{\begin{equation}}
\newcommand{\ee}{\end{equation}}
\newcommand{\bea}{\begin{eqnarray}}
\newcommand{\eea}{\end{eqnarray}}

\def\simlt{\stackrel{<}{{}_\sim}}
\def\simgt{\stackrel{>}{{}_\sim}}
\begin{document}
\title{On the origin of the large scale structures of the universe}

\author{David H. Oaknin}

\affiliation{
Department of Physics and Astronomy, \\
University of British Columbia, Vancouver V6T 1Z1, Canada \\
e-mail: doaknin@physics.ubc.ca}

\begin{abstract}
We revise the statistical properties of the primordial cosmological 
density anisotropies that, at the time of matter radiation equality, 
seeded the gravitational development of large scale structures in the, 
otherwise, homogeneous and isotropic Friedmann-Robertson-Walker flat 
universe. Our analysis shows that random fluctuations of the density
field at the same instant of equality and with comoving wavelength 
shorter than the causal horizon at that time can naturally account, 
when globally constrained to conserve the total mass (energy) of the 
system, for the observed scale invariance of the anisotropies over 
cosmologically large comoving volumes. Statistical systems with 
similar features are generically known as glass-like or lattice-like. 
Obviously, these conclusions conflict with the widely accepted 
understanding of the primordial structures reported in 
the literature, which requires an epoch of 
inflationary cosmology to precede the standard expansion of the 
universe. The origin of the conflict must be found in the 
widespread, but unjustified, claim that scale invariant mass (energy) 
anisotropies at the instant of equality over comoving volumes of 
cosmological size, larger than the causal horizon at the time, 
must be generated by fluctuations in the density field 
with comparably large comoving wavelength.
\end{abstract}

\maketitle

\section{Introduction}

All current cosmological models work within the paradigm in 
which the present distribution of matter in the universe is the result of 
mostly gravitational evolution operating since the time of   
matter-radiation equality on some seed of initial and very small quantum 
fluctuations in the mass (energy) density field of an, otherwise, 
homogeneous, isotropic and flat Friedmann-Robertson-Walker universe.
On the large comoving scales of cosmological interest this seed of 
initial density perturbations at the time of equality is well described by 
the Harrison-Zeldovich spectrum of gaussian fluctuations \cite{HZ}. 

The most characteristic feature of the spectrum of primordial  
anisotropies is the linear dependence, $(\Delta M_V)^2 \propto S$, of the 
variance of mass (energy) fluctuations in restricted spatial sub-volumes 
of comoving cosmological size $L$ on the area $S \sim L^2$ of the 
surface that bounds the sub-volume, rather than on its volume $V \sim 
L^3$. This specific feature $(\Delta M_V)^2 \propto L^2$ is commonly 
known in the literature as scale invariance of primordial mass (energy) 
anisotropies because they produce a gravitational potential $\Phi \sim G 
\frac{\Delta M_V}{L} \sim G$ that does not depend on the comoving scale 
$L$. The scale invariance of the seed of primordial cosmological 
anisotropies at the time of equality was first observationally tested 
ten years ago by the COBE measurements \cite{Bennett:gg} of the 
temperature anisotropies in the cosmic microwave background 
radiation~(CMBR) and has been more recently confirmed by the 
high-precision data of the WMAP collaboration \cite{Spergel:2003cb} and 
others \cite{Netterfield:2001yq}. 

In the context of standard cosmology, nonetheless, comoving scales $L$ of 
cosmological interest, $H^{-1}(t_{eq}) \ll L \simlt H^{-1}(t_0)$, are much 
longer, by orders of magnitude, than the comoving causal scale 
$H^{-1}(t_{eq})$ at the time of matter radiation equality: 
$\frac{H^{-1}(t_0)}{H^{-1}(t_{eq})} \sim 10^4$, where $H^{-1}(t_0)$ is 
the comoving size of the present causal horizon. A central problem in 
contemporary theoretical cosmology is the quantitative explanation of the 
mechanism that produced by the time of equality scale invariant 
anisotropies $(\Delta M_V)^2 \sim L^2$ over cosmological comoving scales 
$L$, which are much larger than the causal horizon at that time. The 
problem arises from the assumption that scale invariant mass (energy) 
anisotropies in comoving volumes of cosmological size $L$ 
must have been produced by fluctuations in Fourier modes of the density 
field with comoving wavelength of comparable size $\lambda \sim L \gg 
H^{-1}(t_{eq})$. We label this assumption as $\label{OSP} [HP]$ for 
easier reference later on. If this assumption were correct, then it would 
be justified to claim that in the context of standard cosmology we lack a 
causal explanation of the mechanism that generated, at $t_{eq}$, scale 
invariant mass anisotropies in comoving volumes much larger than the 
causal horizon at that time. This enigma is usually called the origin of 
structures problem of standard cosmology. The most outstanding of the 
mechanisms that have been invented to {\it solve} this problem is 
inflation, which roughly speaking proposes that the whole observable 
present universe was once causally connected in the remote past and the 
density inhomogeneities were already imprinted at that early time before 
the universe underwent a finite period of exponential expansion which 
stretched it to its current huge size \cite{Turner:2002ts}. This 
mechanism would {\it explain} the presence at the time of equality of 
fluctuation modes in the density field with comoving wavelength 
larger than the standard causal horizon at that instant and with the 
appropriate power spectrum. Obviously, 
the logic of the inflationary explanation of primordial 
structures also relies on assumption [HP].
The main purpose of this work is to bring into attention and revisit the 
assumption $[HP]$ under which the origin of structures problem has been 
raised in the context of standard cosmology.

Assumption [HP] is formally expressed through the fast estimation 
(\ref{berlusco}) of integral (\ref{variance}) for the variance of 
mass (energy) anisotropies over cosmologically large comoving volumes 
(see, for example, eq. 9.1,9.2 and 9.3 in \cite{Mukhanov:1990me}).
According to this estimation mass (energy) anisotropies over volumes $V$ 
of cosmologically large comoving size $L$ are expected to be generated
by the dominant contribution to integral (\ref{variance}) from random 
fluctuations of the stochastic density field (\ref{density}) with 
comparably large comoving wavelength $k \sim 1/L$, while the contributions 
to this integral from the ultraviolet modes $k \gg 1/L$ seemingly cancel 
out and are expected to be sub-dominant or negligible. 
Therefore, under assumption [HP] only a power spectrum with spectral 
index $n = 1$ over the range of cosmologically short comoving momenta 
$k \ll H(t_{eq})$ would be able to generate scale invariant mass 
(energy) anisotropies $(\Delta M_V)^2 \sim L^2$ over cosmologically 
large comoving volumes $L \gg H^{-1}(t_{eq})$, see  
(\ref{preliminaries}). This 
statement is reported in textbooks and research papers in cosmology 
\cite{Peebles}, even though the fast estimation (\ref{berlusco}) is 
since long ago known in statistical mechanics to be incorrect. The 
correct estimation of integral (\ref{variance}) over cosmologically large 
comoving volumes $V$ for a power-law power spectrum with spectral index 
$n$ is given in (\ref{uno}),(\ref{dos}),(\ref{tres}), for all possible 
values of the spectral index. The new estimation shows that any power-law 
spectrum with spectral index $n \ge 1$ would, in fact, generate a pattern 
of scale invariant cosmological mass (energy) anisotropies. 

Scale invariant anisotropies (\ref{tres}) correspond to 
stochastic density fields (\ref{density}) whose random fluctuations have 
been globally constrained to conserve the total mass (energy) of the 
system. The reason why fast estimation (\ref{berlusco}) fails to predict 
the correct behaviour (\ref{tres}) is precisely the fact that in these 
globally constrained systems the largest contribution to integral 
(\ref{variance}) for mass (energy) anisotropies over macroscopically 
large volumes of comoving size $L$ does 
not come from modes with comparably large comoving wavelength $k \sim 
1/L$, whose contribution is suppressed by the global constrain, but from 
the ultraviolet modes $k \simgt H(t_{eq})$ with comoving wavelength 
within the horizon at the time of equality. 
The anisotropies (\ref{tres}) are produced by a mechanism of local 
rearrangement of matter in the uniform universe that can be 
visualized through the following metaphor: people moving short distances 
(i.e. density fluctuations with short comoving wavelength) 
from town to town at different sides of a border can generate 
fluctuations in the total number of inhabitants in each of the countries 
separated by that border (i.e. mass anisotropies over cosmologically 
large volumes), even though the typical size of those countries 
can be much larger than the distance between the two towns. Obviously, 
the longer the common border between the two neighbour countries, the 
larger can be the size of the surface fluctuations in the number of 
inhabitants in each of the countries. Of course, the displacement of 
population does not change the total number of inhabitants in the two 
countries together. This mechanism is well understood in 
statistical mechanics since long ago, but for some reason it has not 
been clearly noticed before in cosmology. Statistical systems that 
show scale invariant anisotropies (\ref{scaleinvariance}) are known in 
the literature as glass-like or lattice-like. In quantum Hall effect 
in condensed matter physics, for example, low-energy excitations are known
to be associated to the borders, as a direct consequence of the global 
constrain imposed by the incompressibility of the gas of electrons 
confined in two spatial dimensions \cite{Halperin}. 

At the light of these comments we realize that all stochastic density 
fields (\ref{density}) characterized by the same average density 
$\rho_0$ and constrained to globally conserve the total mass (energy) of 
the system produce the same pattern of scale invariant mass (energy) 
anisotropies  (\ref{scaleinvariance}) over cosmologically large volumes, 
even though they can be defined by very different power spectra. Hence, 
all power spectra with indexes $n \ge 1$ are, in principle, 
macroscopically indistinguishable. This degeneracy has been ignored for 
many years in the literature, but it is absolutely crucial to understand 
the statistical properties of primordial cosmological mass (energy) 
anisotropies at the time of matter radiation equality.

In next sections we will revisit in detail the arguments and concepts 
that we have briefly introduced in this section and will explicitly show 
that random fluctuations in the density field at the same instant of 
matter radiation equality with comoving wavelength shorter than the 
horizon at that time $H^{-1}(t_{eq})$ can account, when they are 
constrained to conserve the total mass (energy) in the whole system,
for the scale invariant primordial mass (energy) anisotropies over 
cosmologically large comoving volumes.

In the context of this discussion we feel it is necessary for historical 
reasons to comment and clarify a mechanism of local rearrangement of 
matter at the same instant of equality first considered, an discarded, 
long ago by Y.B.~Zeldovich \cite{YaZeldovich} as responsible for the mass 
(energy) anisotropies over cosmologically large comoving volumes. The
apparent incapability of this mechanism to generate the anisotropies is 
cited in textbooks in cosmology \cite{Peebles} as a proof of the 
impossibility in the standard cosmology to generate prior to the time 
of equality scale invariant anisotropies over cosmologically large 
comoving volumes through causally connected physics. The argument is also 
presented as a motivation for the inflationary explanation of 
the origin of primordial structures. We will show, on the contrary, that 
Zeldovich's mechanism does indeed generate scale invariant cosmological 
anisotropies and was discarded only on the ground of a misled analysis 
based on the wrong estimation (\ref{berlusco}).

To model and reanalyze Zeldovich's proposal we consider a free scalar 
hamiltonian density, which is quadratic in the fundamental scalar 
field and its conjugate momentum. Zeldovich nicely noticed that random 
fluctuations of the fundamental fields with comoving wavelength within 
the horizon can, through the quadratic coupling, generate density 
fluctuations with comoving wavelength much larger than the horizon 
$k^{-1} \gg H^{-1}(t_{eq})$. This is a trivial property of Fourier 
analysis. This observation, although correct, was obviously motivated 
by the mistaken assumption [HP]: the 
quadratic coupling offered a causal mechanism to generate at the same 
instant of equality a density power spectrum over the range of 
cosmologically short momenta and, thus, according to fast estimation 
(\ref{berlusco}), mass (energy) anisotropies over cosmologically large 
volumes. On the other hand, a brief and old argument that we 
reproduce in Section V, provided also by Zeldovich, assures that the 
density power spectrum over the range of cosmologically short momenta $k 
\ll H(t_{eq})$ generated through the quadratic coupling of causally 
connected modes must have an spectral index $n \ge 4$ and, therefore, 
it was thought that the variance of mass (energy) anisotropies over 
cosmologically large volumes should decrease, at least, as 
$(\Delta M_V)^2 \sim 1/L$ with the 
comoving size $L$ of the considered volume (\ref{preliminaries}). 
Hence, it was concluded that local rearrangement of matter, although 
able to generate anisotropies over cosmologically large volumes, is 
unable to generate scale invariant cosmological anisotropies $(\Delta 
M_V)^2 \sim L^2$. This conclusion, based on the wrong assumption 
(\ref{berlusco}), is fatally incorrect as we will show: the variance 
of mass (energy) anisotropies generated through local rearrangement 
of matter over cosmologically large volumes are indeed scale 
invariant $(\Delta M_V)^2 \sim L^2$ (\ref{tres}), as we 
should have expected because the fluctuations are globally constrained to 
conserve the total mass (energy) of the system. Zeldovich's 
analysis of his own proposal got wrong because, biased by incorrect 
fast estimation (\ref{berlusco}), it focused on the ability 
to produce a non-zero density power spectrum over the range of 
cosmologically short comoving momenta through the quadratic coupling 
in the hamiltonian density and missed the fact 
that the largest, scale invariant, contribution $(\Delta M_V)^2 \sim L^2$
to cosmological anisotropies comes from random fluctuations of the 
density modes with the shortest wavelength (\ref{tres}), 
while the contribution of density modes 
with cosmologically large comoving wavelength is sub-dominant, 
$(\Delta M_V)^2 \sim 1/L$. In other words, in this model the largest 
cosntribution to mass (energy) cosmological anisotropies does come 
from causally connected random fluctuations of the fundamental fields 
when they couple to produce density fluctuations with comoving 
wavelength still within the horizon, 
and not by modes of the fundamental fields that couple to produce density 
fluctuations with cosmologically large comoving wavelength. 

The irrelevance of the quadratic coupling is evident when, in the same 
framework set up to model Zeldovich's proposal, we consider the 
density field (\ref{density}) to be proportional to the
conjugate momentum field, instead of being given by the quadratic 
hamiltonian density, and study spatial mass (energy) over 
cosmologically large comoving volumes. The momentum field can only 
fluctuate, in this example, in modes with comoving wavelength shorter 
than the horizon and, in absence of quadratic coupling, there is no 
way to generate density fluctuations with wavelength longer than the 
horizon. Yet, an analytical evaluation of integral (\ref{variance}) 
over cosmological volumes show that the anisotropies over volumes of 
comoving size larger than the horizon are still scale invariant.
The reason (\ref{implication}) is the same as above: the conjugate 
momentum field operator is associated to a global charge of the system,
which forces total mass (energy) conservation. In conclusion, it is not 
necessary to generate density fluctuations with cosmologically large 
comoving wavelength in order to produce scale invariant anisotropies over 
cosmologically large volumes. 
Any mechanism of local rearrangement of matter can succesfully 
generate scale invariant anisotropies over cosmologically large volumes.
Zeldovich's old example, properly analyzed, is only a particular 
example.

The analysis that we present here obviously lifts the obstacle that has 
long stood in the way to provide a causal explanation for the origin of 
the large scale structures of the universe in the context of standard 
cosmology (we mean standard cosmological model, without the 
appendix of a preceding inflationary expansion) and provides an attractive 
alternative to inflation to solve the origin of structures problem. In 
principle, this alternative mechanism should probably 
involve only physics at the scale of matter radiation equality, $T_{eq} 
\sim 1~eV$, instead of physics at the very high energy scales up to the 
Planck scale that are usually summoned in inflationary mechanisms. 
Of course, this does not excluse that {\it new} physics could be 
relevant at the instant of equality to explain the origin of primordial 
structures.
Moreover, the generic problem of tuning the initial cosmological 
conditions in inflationary models in order to produce a continued period 
of exponential expansion \cite{Hollands:2002yb} simply vanishes in the 
context of the alternative scenario that we suggest: in this 
alternative scenario it is demanded in a naturally simple way that the 
fluctuating density field is in its ground state at the instant of 
equality.

The paper is organized in nine sections. In Section II we review the 
concepts and tools needed to describe the statistical properties of 
density anisotropies at the time of matter radiation equality in the 
homogeneous, isotropic and flat FRW universe. 
In Section III we characterize the power spectra of gaussian fluctuations 
of the density field that produce scale invariant mass (energy) 
anisotropies over cosmologically large comoving volumes. It is explicitly 
shown that any power-law spectrum ${\cal P}(k) \sim A k^n$ with
spectral index $n \ge 1$ produces an scale invariant spectrum of 
fluctuations $(\Delta M_V)^2 \sim L^2$ in volumes of cosmologically large 
comoving size, $L \simgt H^{-1}(t_{eq})$. This claim is the main reason 
why the analysis performed in this paper disagrees with the main 
conclusions that are found in the literature, where it is claimed 
that only a power-law spectrum with spectral index close to $n = 1$ 
can produce scale invariant mass (energy) anisotropies in 
cosmologically large comoving volumes. The motives of our disagreement 
are extensively discussed in this section III. The calculations presented 
also prove the central claim of this paper, that random fluctuations 
at the same instant of equality in the Fourier 
modes of the density field with comoving wavelength shorter than the 
horizon at that time can account for the observed scale invariant 
primordial anisotropies over cosmologically 
large comoving volumes. We go on in Section IV to discuss the 
characterization of scale invariant mass (energy) random anisotropies in 
terms of the two points correlation function of the fluctuating density 
field. We can advance here that we find that scale invariant anisotropies 
correspond to a certain class of short range correlation functions that 
result when the global mass (energy) contained in the system is 
conserved, and therefore, not allowed to fluctuate. The aim of section V 
is mainly to clarify the arguments that historically laid the so-called 
origin of structures problem [HP] of the standard cosmology. We 
discuss these arguments in the setup of a quantum field theory and 
describe the mechanism of generation of scale invariant primordial 
cosmological mass (energy) anisotropies by local rearrangement of 
matter as random quantum fluctuations of the density field in the 
homogeneous and isotropic ground state of the system at the same 
instant of equality.
Notwithstanding these first five sections of the paper constitute a 
self-contained body, which justifies and proves the claims that appear 
in the abstract, introduction (Section I) and conclusions (Section 
IX) of this paper, we have judged appropriated to include three 
additional sections where we introduce concepts that shall be 
relevant to set up a complete quantum theory of density 
anisotropies coupled to metric perturbations of the homogeneous and 
isotropic FRW space-time background. In Section VI we discuss 
technical concepts on the renormalization of ultraviolet divergent 
expressions in the theoretical setup to conclude
that, in general, the size of the scale invariant spatial anisotropies 
can only be defined as an external renormalizable parameter of the 
theory, like masses or coupling constants. In Section VIII we obtain the
renormalization equations that describe the dependence of this parameter 
on the resolution scale at which the system is probed.
In Section VII we review, at the light of our findings, the linearized 
gauge invariant formalism \cite{Mukhanov:1990me} of density anisotropies 
coupled to metric perturbations of the FRW expanding space-time 
background. Section IX summarizes the conclusions of this paper.

\section{Density anisotropies in homogeneous and isotropic 
statistical systems: concepts and definitions.}

We review in this section some of the concepts and tools of statistical 
mechanics that are widely used to describe the cosmological 
density anisotropies in the early universe at the time $t_{eq}$ of 
matter radiation equality. Our aim is to gather some important results 
from statistical mechanics upon which we will develop our discussion. This 
review will also help us to fix the notation that we use in the rest 
of the paper. For the sake of simplicity we work in comoving coordinates 
and fix the scale factor due to the expansion of the universe to be 
equal to one at the time of equality, $a(t_{eq})=1$.

The statistical system is described by an homogeneous and isotropic
density field $\rho({\vec x})$ in 3D flat space 
$\Omega\equiv {\emph R}^3$:

\begin{equation}
\label{density}
\rho({\vec x}) = \rho_0 + \rho_0 \int
\frac{d^3{\vec k}}{(2\pi)^3}\ {\delta}_{\vec k}\
e^{-i{\vec k}\cdot{\vec x}}.
\end{equation}
If the fluctuations are small the coefficients ${\delta}_{\vec k} = 
\delta^*_{-{\vec k}}$ are independent random complex variables with 
normal distribution, $\langle {\delta}^*_{\vec k_1}\ {\delta}_{\vec k_2} 
\rangle = (2\pi)^3 {\cal P}({\vec k}_1) \delta^3({\vec k_1}-{\vec k_2})$, 
and zero expectation value, $\langle {\delta}_{\vec k} \rangle = 0$, so 
that each local variable $\rho({\vec x})$ is real and has expectation 
value $\langle \rho({\vec x}) \rangle = \rho_0$ independent of the 
position ${\vec x}$. This average value is assumed to be non zero, 
$\rho_0 \neq 0$. The central limit theorem implies that the fluctuations 
of the density field $\rho({\vec x})$ are gaussian. 
The function ${\cal P}({\vec k}) \ge 0$, which 
gives the variance of the fluctuations of each of the random Fourier 
modes $\delta_{\vec k}$, is usually called the power spectrum of the 
statistical fluctuations and it is the most common statistical tool used 
to describe cosmological models. If the system is 
isotropic, the power spectrum ${\cal P}({\vec k})={\cal P}(k)$ depends 
only on the modulus of the momentum that labels each mode, $k\equiv |{\vec 
k}|$. 

Statistical fluctuations of the density field (\ref{density}) in 
macroscopic, but finite, sub-volumes $V \subset \Omega$ are usually 
described in terms of the statistical magnitude $\sigma^2(V) \equiv 
\frac{1}{V^2}\langle 
\left(\int_V d^3{\vec x}\ \frac{\rho({\vec x}) - \rho_0}{\rho_0} 
\right)^2 \rangle$. This magnitude is sometimes called the squared 
density contrast over the volume $V$ and denoted by  
$\left(\frac{\delta\rho}{\rho}\right)_V^2$. 
In terms of the power spectrum of the fluctuations the squared density 
contrast can be expressed as: 

\begin{equation}
\label{sigma2}
\sigma^2(V) = \frac{1}{V^2} \int \frac{d^3{\vec k}}{(2\pi)^3}\ {\cal 
P}(k) |F_V({\vec k})|^2,
\end{equation}
where the geometric factor $F_V({\vec k})$ is given by the expression

\begin{equation}
\label{geometry}
F_V({\vec k}) = \int_V d^3{\vec x}\ e^{-i {\vec k}\cdot{\vec x}}. 
\end{equation}
We wish to notice that this geometric factor $F_V(k)$ is the restricted 
integral over the sub-volume $V$ of the Fourier modes of expansion 
(\ref{density}) in flat 3D space. 

Step by step the proof of relation (\ref{sigma2}) proceeds as follows:
\begin{eqnarray*}
\frac{1}{V^2} \langle \left(\int_V
d^3{\vec x}\ \frac{\rho({\vec x}) - \rho_0}{\rho_0} \right)^2 
\rangle = \frac{1}{V^2} \int_V d^3{\vec x}\ \int_V d^3{\vec y}\ 
\int \frac{d^3{\vec k_1}}{(2\pi)^3}\ 
\int \frac{d^3{\vec k_2}}{(2\pi)^3}\ 
\langle {\delta}^*_{\vec k_1}\ {\delta}_{\vec k_2}\ \rangle
e^{+i {\vec k}_1\cdot{\vec x}} e^{-i {\vec k}_2\cdot{\vec y}} =& \\
\frac{1}{V^2} \int \frac{d^3{\vec k_1}}{(2\pi)^3}\ 
\int d^3{\vec k_2}\ 
{\cal P}(k_1) \delta^3({\vec k_1}-{\vec k_2})
F^*_V({\vec k}_1) F_V({\vec k}_2) = 
\frac{1}{V^2} \int \frac{d^3{\vec k}}{(2\pi)^3}\ 
{\cal P}(k) |F_V({\vec k})|^2. & \\ 
\end{eqnarray*}

The statistical variable $M(V) = \int_V d^3{\vec x}\ \rho({\vec x})$
describes the total mass (energy) contained in the sub-volume $V$.
This is the mathematical object that describes the macroscopic properties 
of the statistical anisotropies in the density field (\ref{density}).
Its average value $\langle M(V) \rangle = \rho_0 V$ is proportional
to the volume $V$ of the considered region. Its variance $(\Delta M_V)^2 
\equiv \langle [M(V) - \langle M(V) \rangle]^2 \rangle = \langle M^2(V) 
\rangle - \langle M(V) \rangle^2 = \langle\left[\int_V d^3{\vec x}\ 
(\rho({\vec x}) - \rho_0) \right]^2\rangle = \sigma^2(V) \times \langle 
M(V) \rangle^2$ measures the typical size of its gaussian fluctuations. 
Combining the previous expressions it is straightforward to obtain:

\begin{equation}
\label{variance}
(\Delta M_V)^2 = \rho^2_0 \int \frac{d^3{\vec k}}{(2\pi)^3}\ {\cal P}(k) 
|F_V({\vec k})|^2.
\end{equation}

The last tool we want to introduce in this section is the two points 
correlation function of the density field, $F({\vec x},{\vec y}) \equiv 
\langle \rho({\vec x})\ \rho({\vec y}) \rangle -\rho^2_0$. If the system 
is homogeneous the two points function must depend only on their relative 
position ${\vec r}={\vec x}-{\vec y}$. If the system is also isotropic 
then the two points function $F(r)$ can depend only on the distance 
between the two points. In terms of the power spectrum, we can write

\begin{equation}
\label{twopoint}
F({\vec r}) = \rho^2_0 \int \frac{d^3{\vec k}}{(2\pi)^3}\ 
{\cal P}(k) e^{+i {\vec k}\cdot{\vec r}}.
\end{equation}
The proof is straightforward:

\begin{eqnarray*}
F({\vec x},{\vec y}) = \rho^2_0 \int \frac{d^3{\vec k_1}}{(2\pi)^3}\ 
\int \frac{d^3{\vec k_2}}{(2\pi)^3}\ 
\langle {\delta}^*_{\vec k_1}\ {\delta}_{\vec k_2}\ \rangle
e^{+i {\vec k}_1\cdot{\vec x}} e^{-i {\vec k}_2\cdot{\vec y}} = & &\\
\rho^2_0 \int \frac{d^3{\vec k_1}}{(2\pi)^3}\ 
\int d^3{\vec k_2}\ 
{\cal P}(k_1) \delta^3({\vec k_1}-{\vec k_2})
e^{+i {\vec k}_1\cdot{\vec x}} e^{-i {\vec k}_2\cdot{\vec y}} 
= & \rho^2_0 \int \frac{d^3{\vec k}}{(2\pi)^3}\ 
{\cal P}(k) e^{+i {\vec k}\cdot({\vec x}-{\vec y})} & 
\end{eqnarray*}
The variance of mass fluctuations (\ref{variance}) can then be 
conveniently expressed in terms of the two point function:

\begin{equation}
\label{variance2function}
(\Delta M_V)^2 = \int_V d^3{\vec x} \int_V d^3{\vec y}\ F({\vec x},{\vec 
y}). \\
\end{equation}
Let us remark at this point that ${\cal P}(k)$ must be integrable over the 
whole 3D momentum space,

\begin{equation}
\label{integrability}
\int d^3{\vec k}\ {\cal P}(k) < \infty,
\end{equation}
if the two points function (\ref{twopoint}) is well defined at ${\vec 
r}=0$ and the squared density contrast (\ref{sigma2}) vanishes when 
integrated over very large volumes, $lim_{V \rightarrow\ \Omega}\ 
\sigma^2(V) = 0$. This condition (\ref{integrability}) allows the 
spectrum to diverge at $k=0$, with the only condition that, if it 
diverges, it does slower than $1/k^3$,

\begin{equation}
\label{conditionZero}
lim_{k \rightarrow 0}\ \left[k^3 \ {\cal P}(k)\right] = 0.
\end{equation}

The discussion in this paper will focus in the statistical properties, 
namely power spectra and two points correlation functions, of 
scale invariant systems, in which the variance of mass (energy) 
fluctuations in restricted sub-volumes $V$ is proportional to the area 
of the surface $S$ that bounds the sub-volume, $(\Delta M_V)^2 \propto 
S$, rather than to its volume $V$. If the considered sub-volume is, in 
particular, an sphere of radius $L$ those statistical systems produce mass 
fluctuations 

\begin{equation}
\label{scaleinvariance}
(\Delta M_V)^2 \propto S \sim L^2.
\end{equation}

Fluctuations in the density field $\rho({\vec x})$ seed in turn 
fluctuations in the gravitational potential through the Poisson's equation 
${\vec \nabla}^2 \Phi(\vec x) = 4\pi G (\rho({\vec x}) - \rho_0)$.
When the mass (energy) fluctuations in cosmologically large comoving 
volumes are scale invariant (\ref{scaleinvariance}) the fluctuations in 
the gravitational potential 

\begin{equation}
\label{potential}
\Delta \Phi_V \propto G \frac{\Delta M_V}{L} \sim G,
\end{equation}
do not depend on the scale $L$.
Statistical systems with this remarkable feature are named after Harrison 
and Zeldovich, who first identified them \cite{HZ}. 

\section{Power spectrum of scale invariant density anisotropies.}

Our aim in this section is to characterize in terms of their power 
spectrum ${\cal P}(k)$ the statistical systems whose mass (energy) 
fluctuations over cosmologically large comoving volumes have scale 
invariant variance (\ref{scaleinvariance}). The two magnitudes are 
directly related by equation (\ref{variance}).

A fast, but incorrect, estimation of relation (\ref{variance}) which, 
nevertheless, appears in many references in the literature 
(see, for example, eq. 9.1,9.2 and 9.3 in \cite{Mukhanov:1990me})
gives:

\begin{eqnarray*}
\label{berlusco}
(\Delta M_V)^2 = \rho^2_0 \int \frac{d^3{\vec k}}{(2\pi)^3}\ {\cal P}(k) 
|F_V({\vec k})|^2 \sim \rho^2_0 \frac{1}{V}\ {\cal P}(k \sim 1/L) 
|F_V(0)|^2 = 
\end{eqnarray*}
\begin{equation}
\label{volumevariance}
\hspace{2.0in} = \rho^2_0 \frac{1}{V}\ {\cal P}(k \sim 1/L) V^2 = 
\rho^2_0\ {\cal P}(k \sim 1/L)\ V,
\end{equation}
where $L$ is the linear size of the considered volume of integration.
This estimation is derived after noticing that the geometric factor 
$F_V(k)$ evaluated at $k=0$ measures the volume $V \sim L^3$ of the 
considered spatial region, $F_V(k=0)=V$, and it decreases fast to zero 
for values of the momentum $k$ larger than the inverse $1/L$ of the 
typical linear size of the considered volume, $F_V(k \simgt 1/L) \sim 0$.

According to this estimation a power-law 
spectrum with spectral index $n$, ${\cal P}(k) \sim A k^n$, over the range 
of comoving momenta $k \simlt H(t_{eq})$ produces mass (energy) 
fluctuations 

\begin{equation}
\label{preliminaries}
(\Delta M_V)^2 \sim \rho^2_0\ {\cal P}(k \sim 1/L)\ L^3 \sim
\rho^2_0\ \frac{A}{L^n}\ L^3 = \rho^2_0\ A\ L^{3-n},
\end{equation}
in cosmologically large comoving volumes, of size $L \simgt H^{-1}(t_{eq})$.
Therefore, a power-law spectrum with spectral index $n = 1$,
${\cal P}(k) \sim A\ k$, over the whole range of comoving momenta $k 
\simlt H(t_{eq})$ seems to be a necessary and sufficient condition to 
generate a pattern (\ref{scaleinvariance}) of scale invariant mass 
(energy) anisotropies over cosmologically large comoving volumes.

Notwithstanding this result (\ref{preliminaries}) is cited in textbooks 
and reviews on the subject \cite{Peebles,Mukhanov:1990me}, we choose to 
check it, both analytically and numerically, integrating equation 
(\ref{variance}). It will turn out that (\ref{preliminaries}) gives a 
reliable estimation only when the spectral index $n$ is in the range, 
$-3 < n < 1$, but the estimation gets fatally wrong for larger values, 
$n \ge 1$. Notice, in particular, that (\ref{preliminaries}) predicts 
an scaling law $(\Delta M_V)^2 \sim L^{\alpha}$, with $\alpha < 2$ for 
values of the spectral index larger than one, $n > 1$. For 
example, for $n = 4$ it predicts $(\Delta M_V)^2 \sim L^{-1}$. Clearly, 
this prediction must be incorrect because it has been rigorously 
proved that $\alpha \ge 2$ \cite{Beck}, the scaling power $\alpha$ is 
necessarily equal or larger than 2 for any homogeneous and isotropic 
statistical system. The fast estimation (\ref{preliminaries}) misses 
that for values of the spectral index $n \ge 1$, in spite of the rapid 
decrease to zero of the geometric factor $|F_V({\vec k})|^2$ for values 
of comoving momenta $k \simgt 1/L$, integral (\ref{variance}) is 
still dominated by the contribution of the very large momenta $k \gg 1/L$, 
rather than by the modes $k \sim 1/L$. This ultraviolet divergence has 
been noticed before in the literature on cosmological 
density anisotropies \cite{Polarski}. Once this contribution is taken 
into account it is found the correct behaviour of the variance of mass 
fluctuations, 
 
\begin{eqnarray}
\label{uno}
\mbox{If} \hspace{0.3in} n \in (-3,0], \hspace{0.3in} (\Delta M_V)^ 2 
\propto & V^{1-n/3} & \sim 
L^{3-n}, \\
\label{dos} 
\mbox{If} \hspace{0.3in} n \in (0,+1], \hspace{0.3in} (\Delta M_V)^ 2 
\propto & S^{3/2-n/2} & \sim 
L^{3-n}, \\
\label{tres} 
\mbox{If} \hspace{0.3in} n \in [+1,\infty), \hspace{0.3in} (\Delta M_V)^ 
2 
\propto & S & \sim 
L^{2}.
\end{eqnarray}

These correct results (\ref{uno},\ref{dos},\ref{tres}), obviously, respect 
the bound $\alpha \ge 2$. We wish to remark that these results are not
new neither unexpected. They were previously reported, for 
example, in \cite{Gabrielli:2001xw} in the context of a discussion of the 
spectrum of primordial cosmological anisotropies. They are also very 
well-known in condensed matter physics in the description of 
glass-like systems. We rederive them here only with the aim of convincing 
the reader about their validity. As we will show, the disagreement between 
the old estimation (\ref{preliminaries}) and the correct result  
(\ref{tres}) listed above is the only reason why the conclusions of this 
paper do not agree with those widely accepted and reported in the 
literature.  

In order to justify the result (\ref{tres}) we now consider as 
volume $V$ of integration an sphere of comoving radius $L$. The 
rotationally symmetric geometry of the considered 
spherical sub-volume does not change the qualitative features of the 
results that we want to prove, but it allows to obtain the geometric 
factor (\ref{geometry}) analytically:

\begin{equation}
F_V(k) = 4\pi L^3 \frac{1}{(k L)^3}
\left(sin(k L) - (k L)\hspace{0.03in} cos(k L) \right).
\end{equation}
When this expression is introduced in (\ref{variance}) we obtain the 
relation:

\begin{equation}
\label{integral}
(\Delta M_V)^2 = 8 \rho^2_0\ L^3 \int d(k L)
\frac{{\cal P}(k)}{(k L)^4} \left(sin(k L) - (k L)\hspace{0.03in} cos(k 
L) \right)^2.
\end{equation}
It is important to notice that ${\cal I}(w) = \frac{1}{w^4} 
\left(sin(w) - w\ cos(w) \right)^2$, where $w=k L$, is not singular
at $w=0$ as the factor $\frac{1}{w^4}$ could induce to think, because
$(sin(w) - w\ cos(w))^2 = {\cal O}(w^6)$ when $w \rightarrow 0$.

For the power spectrum we now assume a power-law dependence ${\cal P}(k) = 
{\cal A} \frac{k^n}{k^{n+3}_c}$ over a large, but finite, range of 
comoving momenta $0 \le k \simlt k_c$, and beyond which the power 
spectrum is cutoff by some supression factor. Later on we will specify 
the scale $k_c$ of the cutoff procedure and will characterize it as a 
regularization step of an ultraviolet divergent expression in the context 
of the renormalization programme of a physical parameter. For the moment 
it is enough to say that this scale can be somewhat of the order of the 
Hubble horizon at the time of equality, $k_c \simgt H(t_{eq})$. Notice 
also that coefficient ${\cal A}$ is dimensionless and can be fixed 
independently of the spectral index $n$. For the sake of simplicity we 
will assume in the analytical proof that we present below that the cutoff 
at $k_c$ in the power spectrum is sharp, ${\cal P}(k \ge k_c) = 0$. 
More realistic exponential or polynomial cutoffs are evaluated 
numerically. The results are presented in Fig. 1. They show the same 
behaviour (\ref{uno},\ref{dos},\ref{tres}) that we observe from the 
analytical 
discussion of the simpler case with a sharp cutoff. 

\begin{figure}
\begin{flushleft}
{\epsfig{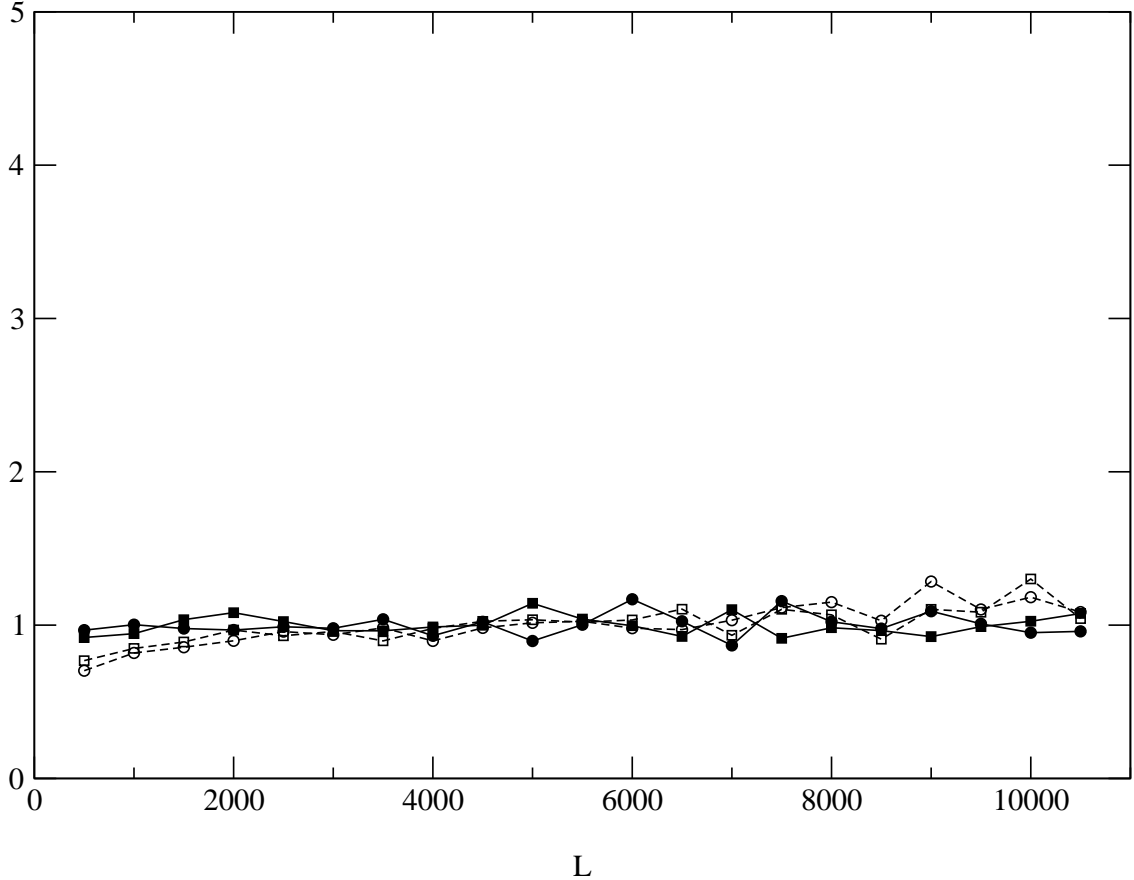}}
\end{flushleft}
\caption{The points on the plot represent the numerically evaluated 
normalized ratio $\frac{(\Delta M_L)^2}{L^2}$ of the variance of mass 
(energy) fluctuations in spherical 3D volumes of comoving 
radius $L$ to the area of its surface $S=4\pi L^2$. Two different 
power-law spectra ${\cal P}(k) = {\it a} k^n$ with positive spectral 
indices are considered: $n = 1$ (dashed lines) and $n = 4$ (straight 
lines). The normalized data clearly show that in both cases the ratio 
remains constant over a very large range $\left(100 \longleftrightarrow 
10000\right)$ of values of the comoving radius $L$, measured in 
arbitrary units of length. The cutoff scale has been 
chosen arbitrarily at $k^{-1}_c \simeq 0.1$ units of length, so that
the dimenssionless parameter $k_c L \gg 1$ ranges from $10^3$ to $10^5$.
Two different setups to cutoff the power spectrum in the ultraviolet 
modes has been considered in this graph: an exponential cutoff ${\cal 
P}(k) \mbox{Exp}(-k/k_c)$ (squares); and a polynomial cutoff ${\cal P}(k) 
\left(\frac{1}{1+(k/k_c)^4}\right)$ (circles).} \label{F1}
\end{figure}

We now discuss separately the four different cases when the spectral 
index $n \in {\emph R}$ in the power-law is: case I) $n > 1$; case 
II) $n = 1$; case III) $0 < n < 1$; and, finally, case IV) $-3 < n \le 0$.
The spectral index $n \in R$ is restricted by condition  
(\ref{conditionZero}) to be larger than $n > -3$.

In case I), $n > 1$, the variance (\ref{integral}) can be estimated 
with very good accuracy by the analytical expression

\begin{equation}
\label{integral2}
(\Delta M_V)^2 \sim 4 \rho^2_0\ L^3 \int_{0}^{k_c L} 
d(k L) \frac{{\cal P}(k)}{(k L)^2},
\end{equation}
if $ L \gg k^{-1}_c$. The reader can visualize the condition
for this estimation after noticing that the factor 
$\frac{1}{(k L)^2} \left(sin(k L) - (k L)\hspace{0.03in} cos(k L) 
\right)^2$ in the integrand of (\ref{integral}) approximately halves the 
area left under the curve $\frac{{\cal P}(k)}{(k L)^2}$. When the last 
integral is performed we obtain

\begin{equation}
\label{integral3}
(\Delta M_V)^2 \sim 4 \rho^2_0\ L^3 \int_{0}^{k_c L} 
d(k L) \frac{{\cal A} (k^n/k^{n+3}_c)}{(k L)^2} = 4 \rho^2_0\
\frac{\cal A}{(n-1)} \frac{1}{k^4_c} L^2 \equiv \rho^2_0\
{\cal A'} \frac{1}{k^4_c} L^2,
\end{equation}
where ${\cal A'}$ is a dimensionless factor that fixes the 
absolute amplitude of the fluctuations. Notice the promised linear 
dependence of the variance on the area $S \propto L^2$ of the surface 
that bounds the sphere of integration. It is also interesting to 
realize that integral (\ref{integral}) is convergent at $k L \rightarrow 
0$, but it is divergent in the modes with very large momenta $k \sim 
k_c \gg 1/L$ close to the cutoff. The explicit way how the cutoff in 
${\cal P}(k)$ regularize this expression is not qualitatively relevant in 
this discussion, as it is shown in Fig. 1, but the need for  
regularization proves that the largest contribution to the variance comes 
from fluctuations of the Fourier modes $\delta_k$ of the density field 
(\ref{density}) in the ultraviolet range $k \sim k_c$. We also can 
understand this result by noticing that, according to 
(\ref{preliminaries}), the contribution to integral (\ref{variance}) over 
comoving volumes of cosmological size $L$ coming from modes with 
comparable wavelength $k \sim 1/L$ decreases for $n > 1$ as $\sim L^{3 - 
n}$, which is subdominant to the contribution (\ref{tres}) $\sim L^2$ from 
the ultraviolet modes close to the cutoff $k \sim k_c$.

It is also important to notice how the ultraviolet cutoff $k_c$ enters 
expression (\ref{integral3}): this regularization cutoff only modifies the 
absolute size of the mass (energy) fluctuations, but it does not modify 
the scale invariant linear dependence of their variance on the area 
$S$ of the surface that bounds the considered sub-volume 
(\ref{scaleinvariance}). We will discuss this point in more detail in 
next Section VI and Section VIII. 

Before going on to cases II, III) and IV) let us further clarify the 
result (\ref{integral3}) we have just proved. We have shown that 
fluctuations in the modes $\delta_k$ of the density field (\ref{density}) 
with the shortest comoving wavelength $k \simgt  
H(t_{eq})$, can generate scale invariant mass (energy) anisotropies over 
much larger volumes of comoving cosmological size $L \gg H^{-1}(t_{eq})$. 
We have been asked where is the 'non-linear' mechanism that leads to this 
result ? The answer is in the integral expression (\ref{variance}) for 
the variance of mass anisotropies: the reader can clearly see 
that fluctuations in all the modes 
$\delta_k$ over the whole range $0 \le k \le k_c$ add up their
positive contribution $dk\ k^2\ {\cal P}(k)\ |F_{V}({\vec k})|^2 > 0$ to 
the total variance. It is commonly assumed that for a given volume $V$ of 
cosmological comoving size $L \gg k_c^{-1}$, the overwhelmingly largest 
contribution to this integral (\ref{variance}) comes from the modes $k 
\sim L^{-1}$ (see, for example, eq. (9.3) in \cite{Mukhanov:1990me}). This 
is the assumption [HP], which leads to the 'linear' relationship 
(\ref{berlusco}) between the variance of mass (energy) anisotropies in 
a volume of comoving size $L$ and fluctuations ${\cal P}(k \sim 
L^{-1})$ of the modes $\delta_k$ with comparable comoving wavelength 
$L \sim k^{-1}$. We have explicitly shown above that this 'linear' 
relationship does not hold in case I), when fluctuations of the the 
ultraviolet modes $k \sim k_c$ produce the largest contribution to the 
global integral (\ref{variance}) and generate 
scale invariant mass (energy) anisotropies. The 'non-linear' connection 
between mass (energy) anisotropies in cosmologically large comoving 
volumes and fluctuations in the Fourier modes of the density field with 
much shorter comoving wavelength is intrinsically wired into the integral 
expression (\ref{variance}), which stands as the core hardware 
of the formal description of the macroscopic properties of gaussian 
fluctuations in the density field (\ref{density}). This 
'non-linear' connection is relevant in case I), when spatial 
anisotropies are scale invariant, but has been carelessly disregarded in 
the literature on the origin of cosmological primordial density 
anisotropies. 

In case II), $n = 1$, the integral (\ref{integral2}) is apparently 
divergent not only at $k L \gg 1$ but also at $k L \rightarrow 0$. Now it 
is important to remember that, as ${\cal P}(0)=0$ for any $n > 0$, 
the integrand ${\cal P}(k) {\cal I}(k L)$ in (\ref{integral}), of which 
(\ref{integral2}) is only an estimate, is regular at $k L \rightarrow 0$.
The variance is still dominated by the ultraviolet modes $k \sim k_c \gg 
1/L$, but now they contribute only a sub-leading logarithmic divergence,

\begin{equation}
(\Delta M_V)^2 \sim 4 \rho^2_0 {\cal A}\ \frac{1}{k^4_c} 
\hspace{0.03in} L^2\ ln(k_c L).
\end{equation}
To leading order the variance $(\Delta M_V)^2 \sim L^2$ shows 
the same linear dependence on the area of the boundary surface that was 
obtained for the variance in case I).

In case III), $0 < n < 1$, the integrand in 
(\ref{integral}) decreases fast enough in the large $k$ modes to 
render the integral over these modes convergent. A proper estimation 
shows that now the result agrees with the estimation in 
(\ref{preliminaries}), $(\Delta M_V)^2 \sim \rho^2_0 {\cal B'}\ 
\frac{1}{k^{3+n}_c} \hspace{0.03in} L^{3-n}$, because the integral 
now is dominated by the 'linear' modes $k_c L \sim 1$. 

Finally, we consider the case of negative values of the spectral 
index, $-3 < n \le 0$. In this case, the integrand in the 
l.h.s of (\ref{integral}) decreases even faster to zero in the ultraviolet 
regime $k \sim k_c \gg 1/L$ and the integral is dominated by the modes 
$k L \sim 1$. The fast estimation presented in (\ref{preliminaries}) is, 
therefore, also valid in this case. In fact we obtain,

\begin{equation}
(\Delta M_V)^2 \sim \rho^2_0 {\cal B''} \frac{1}{k^{n+3}_c} 
L^{3-n}. 
\end{equation}

This completes our proof of (\ref{uno},\ref{dos},\ref{tres}). These 
results show that any power-law spectrum over a finite range of comoving 
momenta $0 \le k \simlt k_c$, that vanishes at $k = 0$, i.e. ${\cal 
P}(0)=0$, and whose first derivative at the origin also vanishes, i.e. 
${\cal P}'(0)=0$, produces a pattern of scale invariant mass (energy) 
fluctuations in any volume whose comoving size is much larger 
than the comoving length scale of the cutoff, $k_c L \gg 1$: 

\begin{equation}
\label{equivalence1}
(\Delta M_V)^2 \propto S \sim L^2 \hspace{0.3in} \Leftrightarrow 
\hspace{0.3in} \left({\cal P}(0)=0 \hspace{0.3in} \vee \hspace{0.3in} 
{\cal P}'(0)=0 \right). \\
\end{equation}
 
If only the power spectrum vanishes at zero, ${\cal P}(0)=0$, but the 
first derivative does not, ${\cal P}'(0) \neq 0$, we still have surface
dependance (\ref{dos}) for the variance of mass (energy) anisotropies, 
$(\Delta M_V)^2 \propto S^{\gamma}$, although the polynomial growth is 
faster than linear $1 < \gamma \le 1.5$. On the other hand any power 
spectrum which does not vanish at $k = 0$, i.e ${\cal P}(0) \neq 0$,  
produces in spatial regions whose typical comoving size is 
much larger than the comoving length scale of the cutoff $k_c L \gg 1$
a pattern of mass (energy) anisotropies $(\Delta M_V)^2 \propto 
V^{\beta}$, with $1 \le \beta < 2$, proportional to some power of the 
volume. We will further comment on this point in next section. Now we do 
not want to derail from the main discussion of this section that shall 
focus on power-law spectra with spectral index $n > 1$ or any linear 
combination of them, which produce strict scale invariant anisotropies 
(\ref{scaleinvariance}) over cosmologically large comoving volumes through 
the dominant contribution of the Fourier modes with comoving wavelength 
shorter than the horizon $k \simgt H(t_{eq})$.

The result that we report in equation (\ref{tres}) has further 
implications that we want to emphasize:
power spectra with different spectral indices $n > 1$ produce gaussian
fluctuations in the macroscopic variable $M(V) \equiv \int_V
\rho({\vec x})$ with the same average value $\rho_0 V$ and the same
characteristic variance $(\Delta M_V)^2 = \rho^2_0 {\cal A'} 
k^{-4}_c L^2$ in any macroscopic volume $V$, for their amplitude can 
independently be fixed through the dimensionless prefactor ${\cal A'}$. 
Therefore, we must conclude that power spectra with different spectral 
indices $n > 1$, or linear combinations of them, are macroscopically 
indistinguishable in the scale invariant mass (energy) anisotropies 
$\Delta M_V$ that they produce in cosmologically large spatial 
sub-volumes. 

If all different power spectra ${\cal P}(k) \sim k^n$, with index $n >
1$, produce mass (energy) anisotropies in macroscopic volumes with the 
same variance, which scales linearly with the boundary area 
(\ref{scaleinvariance}), all of them produce scale invariant 
gravitational potentials (\ref{potential}). Hence, there is no way to 
formally distinguish between power-law spectra with different $n > 1$ 
through the macroscopic mass fluctuations they produce, neither through 
their gravitational effect on large macroscopic distances. 

Does this discussion mean that power spectra with different spectral index
$n > 1$ are physically indistinguishable ? Not exactly, in principle. 
Different power spectra would produce different two points correlation 
functions $F({\vec r})$, but the differences are hidden when the 
macroscopic magnitudes $(\Delta M_V)^2$, given by 
(\ref{variance2function}), are compared. We will come back to this point 
later in Section VI and Section VIII.

We finish this section with an estimation of the density contrast 
$(\delta \rho/\rho)_L \equiv \sqrt{\sigma^2_L}$ in spherical 
volumes of comoving size $L$ associated to scale 
invariant mass (energy) anisotropies (\ref{integral3}). We have:

\begin{equation}
\sigma^2_L = \frac{(\Delta M_V)^2}{\langle M_V \rangle^2} \sim
\frac{\rho^2_0\ {\cal A'} \frac{1}{k^4_c} L^2}{\rho^2_0 V^2} = 
\frac{\cal A'}{(4\pi/3)^2} \frac{1}{(k_c L)^4}.
\end{equation}

Therefore, $(\delta \rho/\rho)_L \sim \frac{\sqrt{\cal A'}}{(4\pi/3)}
\frac{1}{(k_c L)^2} \sim 0.25\ \sqrt{\cal A'} \frac{1}{(k_c L)^2}$.
Present bounds on the density contrast over cosmological scales
constrain $(\delta \rho/\rho) \simlt 10^{-5}$. Therefore, 
we can constrain the scale $k_c$ of momentum cutoff to be such that
$k_c L \simgt 100\ ({\cal A'})^{1/4}$ for all cosmologically large 
comoving scales, $H(t_{eq}) L \gg 1$. Both expressions together demand 
that ${\cal A'} < \left(k_c/H(t_{eq})\right)^4$, so that $k_c$ can be 
naturally larger than $H(t_{eq})$.

\section{Scale invariant density anisotropies and total mass 
conservation.}

We want to obtain in this section the conditions that characterize the two 
points correlation function (\ref{twopoint}) of statistical systems which 
show scale invariant mass (energy) fluctuations (\ref{scaleinvariance}) 
over macroscopically large, but finite, spatial sub-volumes $V \subset 
\Omega$ of the whole system. These conditions will be then used to prove 
that random fluctuations of the density field (\ref{density}) produce 
scale invariant mass (energy) anisotropies, $(\Delta M_V)^2 \propto S$, 
in cosmologically large comoving volumes if an additional constrain of 
total mass (energy) conservation is imposed:

\begin{equation}
\label{implication}
(\Delta M_{\Omega})^2 = 0 \hspace{0.3in} \Rightarrow \hspace{0.3in} 
(\Delta M_{V})^2 \propto S.
\end{equation} 

Let us start then characterizing the two points function of scale 
invariant density fluctuations. Equation (\ref{twopoint}) can be inverted 
and written as

\begin{equation}
\label{expression}
{\cal P}({\vec k}) = \frac{1}{\rho^2_0} \int_{\Omega} d^3{\vec r}\
e^{-i {\vec k}\cdot{\vec r}} F({\vec r}) =
\frac{4\pi}{\rho^2_0} \int_0^{\infty} dr\ r^2 
\frac{sin(kr)}{kr} F(r).
\end{equation}

The power spectrum is by definition a positive function, ${\cal P}(k) \ge 
0$. This condition constrains any possible choice for the two 
points correlation function $F(r)$. Yet the most interesting constrain 
is derived from the condition that appears in the right hand side of 
the logical equivalence (\ref{equivalence1}): ${\cal P}(k)_{k=0} = 0$ and
$\frac{d{\cal P}(k)}{dk}|_{k=0} = 0$. 

First, we notice 

\begin{equation}
\label{equivalence4}
{\cal P}(0) = 0 \hspace{0.3in} \Leftrightarrow \hspace{0.3in} 
\frac{1}{\rho^2_0} \int_0^{\infty} dr\ r^2 F(r) = 0.
\end{equation} 
This is the only non-trivial constrain that we need to request on the two 
points correlation function $F(r)$ in order to get scale invariance mass 
(energy) anisotropies according to (\ref{equivalence1}), because the 
condition on the first derivative at the origin $k = 0$ is trivially 
fulfilled if the linear operators $\frac{d}{dk}$ and $\int_0^{\infty} dr$ 
commute when acting on the function $g(k,r) = r^2 \frac{sin(kr)}{kr} 
F(r)$:

\begin{equation}
\label{expressionprime}
{\cal P}'(0) = 
\frac{4\pi}{\rho^2_0} \frac{d}{dk}\left(\int_0^{\infty} 
dr\ r^2 \frac{sin(kr)}{kr} F(r)\right)_{k=0} = 
\frac{4\pi}{\rho^2_0} \int_0^{\infty} 
dr\ r^2 \frac{d}{dk}\left(\frac{sin(kr)}{kr}\right)_{k=0} 
F(r) = 0,
\end{equation}
We will not discuss here the formal mathematical requirements needed to 
get this technical condition (\ref{expressionprime}) satisfied and 
will assume, for simplicity, that it is trivially fulfilled in the 
setup we have defined. In fact, in all the standard definitions 
introduced in Section II it was assumed that linear operators in real 
${\vec r}$-space and momentum ${\vec k}$-space do commute. Of course,
a further consideration of these formal aspects would be interesting.

The logical equivalence (\ref{equivalence1}), therefore, implies that 
condition $\frac{1}{\rho^2_0} \int_{\Omega} d{\vec r} F(r) = 0$  
in the r.h.s. of (\ref{equivalence4}) completely characterizes the two 
points function $F(r)$ of statistical systems which produce scale 
invariant mass (energy) anisotropies. This condition on the two points 
function is known in statistical mechanics to characterize 
glass-like systems. In quantum Hall effect in condensed matter physics 
a similar constrain on low-energy excitations is associated to the 
incompressibility of the gas of electrons confined in 2D 
spatial dimensions. These excitations are known to be 
associated to the borders \cite{Halperin}. 

The condition above can be written in a different way as 

\begin{eqnarray*}
\frac{1}{\rho^2_0} \int_{\Omega} d{\vec r} F(r) = 
\frac{1}{\rho^2_0} \int_{\Omega} d{\vec r} \left(\langle \rho({\vec x})\
\rho({\vec x} + {\vec r}) \rangle - \langle \rho({\vec x}) \rangle \langle
\rho({\vec x} + {\vec r}) \rangle \right) = \\
= \frac{1}{\rho^2_0} \left(\langle \rho({\vec x}) M(\Omega) \rangle 
- \langle \rho({\vec x}) \rangle \langle M(\Omega) \rangle \right) = 0,
\end{eqnarray*}
where $M(\Omega) = \int_{\Omega} d{\vec x} \rho(\vec x)$ is the total mass 
in the whole space $\Omega$. The condition of scale invariance 
(\ref{equivalence4}), therefore, is satisfied if and only if the total 
mass (energy) $M(\Omega)$ is not correlated to the specific value of the 
density field at any point ${\vec x}$:

\begin{equation}
\label{mass}
\langle \rho({\vec x}) M(\Omega) \rangle
= \langle \rho({\vec x}) \rangle \langle M(\Omega) \rangle.
\end{equation}
Furthermore, if we integrate the previous expression over a finite volume
$V$ we obtain that in systems that show scale invariant
random mass (energy) anisotropies (\ref{scaleinvariance}),

\begin{equation}
\label{massV}
\langle M(V) M(\Omega) \rangle
= \langle M(V) \rangle \langle M(\Omega) \rangle,
\end{equation}
the total mass (energy) contained in the system, $M(\Omega)$, is not 
correlated to the mass (energy) contained in any finite sub-volume 
$V \subset \Omega$ inside it.
Of course this condition is easily fulfilled if the total mass $M(\Omega)$
is constrained to a constant value and cannot fluctuate. Hence, we 
conclude that, in the context of standard cosmology (again, without 
inflation) random fluctuations of the density field (\ref{density}) 
are necessarily scale invariant (\ref{scaleinvariance}) if the total mass 
of the system is constrained and it is not allowed to fluctuate 
\cite{Oaknin:2003dc}. This statement is expressed through the logical 
implication (\ref{implication}).

The surface dependence of the variance of mass (energy) fluctuations
in a system whose global energy does not fluctuate is easily 
understandable: mass (energy) fluctuations in any two complementary 
closed sub-volumes $V$ and ${\widetilde V} = cl(\Omega - V)$ must be 
correlated, such that $(\Delta M_V)^2 = (\Delta M_{\widetilde V})^2$, 
because $M_{\Omega} = M_V + M_{\widetilde V}$ is an extensive magnitude. 
This correlation implies that the variance of mass fluctuations in 
sub-volume $V$ must be only a function $(\Delta M_V)^2 = f(S)$ of the 
area of its common boundary with ${\widetilde V}$. This is what we found
in (\ref{dos}) and (\ref{tres}): when $n > 0$ global mass (energy)
condition (\ref{massV}) is satisfied because ${\cal P}(0) = 0$.  
Furthermore, technical condition (\ref{expressionprime}) guarantees 
the linear dependence (\ref{tres}) once this condition 
(\ref{massV}) is fulfilled.

\section{Cosmological density anisotropies from short wavelength 
rearrangements in the distribution of matter.}

At this point of our discussion we feel necessary to revisit and 
comment an old argument first discussed by Y.B.~Zeldovich 
\cite{YaZeldovich} and, after him, by many other authors in textbooks 
\cite{Peebles} and research papers \cite{Carr}. The argument 
discusses, under assumption [HP], the possibility that causally 
connected processes at the instant of matter radiation equality could 
account in the standard cosmology (again, without inflation) for the 
cosmological mass (energy) anisotropies at that time.

According to mistaken estimation (\ref{berlusco}), derived from assumption 
[HP], the mechanism responsible for scale invariant mass (energy) 
anisotropies over cosmologically large comoving volumes at the instant of 
equality would need to generate a density power spectrum ${\cal P}(k) 
\sim k$ over the range of cosmologically short comoving momenta. That is, 
the mechanism would need to justify: first, the presence at the instant 
of equality of fluctuating modes in the density field with comoving 
wavelength much larger than the causal horizon at that time; second, the 
linear growth with the momentum scale of the power stored in these modes. 
The first obstacle seems easily superable: trivial Fourier analysis is 
enough to notice that fluctuations at the same instant of equality in 
modes of weakly interacting fields with comoving wavelength shorter than 
the horizon can couple through the quadratic kinetic terms of the 
hamiltonian density to produce density fluctuations with cosmologically 
large comoving wavelength.

On the other hand, the second requirement seems unattainable. The power 
spectrum that quadratic coupling creates over the range of cosmologically 
short comoving momenta, $k \ll H(t_{eq})$, is not linear, but quartic, 
in the momentum of the Fourier modes: ${\cal P}(k) \sim k^4$. The 
explicit calculation can be found in the textbooks. We reproduce it 
below in this section. For now, the following fast estimation will 
suffice:

\begin{eqnarray}
\label{assessment}
\nonumber
&\delta_k & \simeq 
\int_V d^3{\vec x}\ \frac{\rho({\vec x}) - 
\rho_0}{\rho_0}\ e^{+i{\vec k}\cdot {\vec x}} \simeq \\ & & \simeq 
\int_V d^3{\vec x}\ \frac{\rho({\vec x}) - 
\rho_0}{\rho_0} + i \int_V d^3{\vec x}\ \left({\vec k}\cdot {\vec 
x}\right)\ \frac{\rho({\vec x}) - \rho_0}{\rho_0} + O(k^2).
\end{eqnarray}  
If the total mass is conserved, then the first, $k$-independent, term of 
the expansion cancels out, which implies $\delta_k \sim O(k)$ that the 
power spectrum of the fluctuations ${\cal P}(k) = |\delta_{k}|^2 \simeq 
O(k^2)$ and, therefore, ${\cal P}(0)=0$. This is in full agreement with 
our results (\ref{implication}) and (\ref{equivalence1}) of the previous 
section. In addition it is also noticed in the literature that the linear 
term in ${\vec k}$, second in the above expansion, must also vanishes 
if the center of mass of the distribution does not fluctuate. In other 
words, conservation laws of the mass distribution demand $\delta_k 
\sim O(k^2)$. In consequence, the power spectrum of mass (energy) 
fluctuations generated locally grows, at least, as the quartic power 
$n = 4$ of the momentum, ${\cal P}(k) = |\delta_k|^2 \sim k^4$, over the 
whole range of cosmologically short momenta, $k \simlt H(t_{eq})$.

Introducing this quartic power spectrum in estimation 
(\ref{preliminaries}), derived from assumption [HP] through eq. 
(\ref{berlusco}), we obtain the clear 
prediction $(\Delta M_V)^2 \sim 1/L$, that the variance of mass (energy) 
cosmological anisotropies generated by local modes of a fundamental field 
that couple quadratically in the density field seem to decrease at least 
inversely proportional to the comoving size of the considered volume.
On the ground of this analysis it is usually claimed in cosmology the 
impossibility to causally generate in the standard cosmology scale 
invariant mass (energy) anisotropies $(\Delta M_V)^2 \sim L^2$ over 
cosmologically large comoving volumes. This conclusion is known as the 
origin of structures problem of standard cosmology. We wish the reader 
to realize how heavily this conclussion relies on estimation 
(\ref{berlusco}),(\ref{preliminaries}), which was derived from 
assumption [HP].

Let us now show the inconsistency of the previous analysis, highlighting 
the use of assumption [HP] and the assessment (\ref{assessment})
for the power spectrum in Zeldovich's proposal, ${\cal P}(k) \sim k^4$ 
over the range of cosmologically short momenta. We have learned in 
Section III, case I, that estimation [HP], written down 
in eq. (\ref{preliminaries}), for mass (energy) anisotropies with a 
quartic, $n = 4$, power spectrum ${\cal P}(k) \sim k^4$, gets absolutely 
wrong because it does not properly take into account the contribution 
from density fluctuations with very short comoving wavelength. In fact, 
according to the correct estimation (\ref{tres}) the anisotropies 
generated by the power spectrum with spectral index $n = 4$ 
(\ref{assessment}) are indeed scale invariant, as we should have expected 
since the very beginning of this section from the global constrain on 
total mass (energy) conservation (\ref{implication}).
 
Zeldovich's proposal is only a particular example of a more general 
mechanism of generating scale invariant mass (energy) anisotropies 
over cosmologically large comoving volumes through the causally 
connected local rearrangement of matter at the time of equality in a 
previously uniform universe. How does this mechanism work?
Displacement of masses through the border that bounds the considered 
sub-volume changes the total masses contained both inside and outside 
that volume, although the total mass (the sum of the mass inside plus the 
mass outside) is conserved and, therefore, the cosmological mass 
anisotropies it generates must be scale invariant (\ref{implication}).

This mechanism works even when the density 
field is considered to be linear, instead of quadratic, in the 
fundamental fields, as we explicitly show below in this section. In fact, 
the capability in Zeldovich's original proposal to generate density 
fluctuations with cosmologically large comoving wavelength through the
quadratic coupling of fundamental fields is only a secondary feature
without relevance, because the contribution of these very long modes to 
integral (\ref{variance}) for the variance of mass (energy) cosmological 
anisotropies is negligible compared to the contribution of modes of the 
density field with short comoving wavelength within the horizon.

In order to make these arguments explicit and absolutely clear we want to 
show how Zeldovich's mechanism works in a toy-model and point out 
where the analysis of the mechanism gets wrong in the literature, under 
assumption (\ref{berlusco}). We consider a massless and free 
quantum scalar field $\phi(x_{\nu})$ in the homogeneous, isotropic and 
flat FRW space-time background $ds^2 = a^2(x_0)\left(dx^2_0 - 
\sum_{i=1,2,3.} dx^2_i\right)$,

\begin{equation}
\label{action}
{\cal S} = \int d^4x^{\nu}\ a(x^0)
\partial_{\mu}\phi(x^{\nu})^{\dagger}
\partial^{\mu}\phi(x^{\nu}), 
\end{equation} 
where $x^{\nu}$ are comoving coordinates and $a(x_0)$ is the scale 
factor of the expansion, which we fixed to be equal to one at the 
instant of equality, $a(x^0_{eq})=1$. Let us state clearly that the 
scalar field $\phi$ we use here is not the inflaton, neither it is any 
new particle that we are inventing. Action (\ref{action}) is simply a 
toy-model of the mechanism proposed and discussed by Y.B.~Zeldovich long 
ago. We could have chosen to show this mechanism in a system of fermions 
or gauge bosons.

We are interested in describing the spectrum of energy (mass) 
anisotropies over cosmologically large comoving volumes due to local 
displacements of matter within the causal horizon at equality, which 
we describe as quantum fluctuations in the Fourier modes of the scalar 
field 

\begin{equation}
\label{field}
\phi({\vec x}) = \int_{k_{IR}} \frac{d^3{\vec k}}{(2\pi)^3}
\frac{1}{\sqrt{|{\vec k}|}} Q_{\vec k} e^{+i{\vec k}\cdot{\vec x}}
\end{equation}
with comoving wavelength shorter than the causal horizon, $k^{-1} \simlt 
H^{-1}(t_{eq})$. Therefore, in our toy-model we introduce an infrared 
cutoff in momentum $k_{IR} \simeq H(t_{eq})$ in the phase space of the 
scalar quantum field. The way how this infrared cutoff is introduced is 
not relevant, it can be a sharp cutoff or a mass term in the action 
(\ref{action}) of the scalar field. The important point is that we want to 
center the attention on the role of the under-horizon modes at the time of 
equality in the generation of mass (energy) anisotropies over 
cosmologically large comoving volumes. In Section VII this toy-model is 
embedded into a physically motivated description of the anisotropies, the 
linearized gauge invariant formalism \cite{Mukhanov:1990me}. There we 
will comment on the role of super-horizon modes of the fluctuating field 
(\ref{field}).

The dynamics of Fourier modes of (\ref{field}) with comoving wavelength 
within the horizon is not affected by the expansion of the universe and 
can be described by a free hamiltonian in flat Minkowski space-time,

\begin{equation}
\label{hamiltonian}
{\cal H} = \int_{\Omega} d^3{\vec x} \hspace{0.07in} {\it h}({\vec x}) =
\int_{\Omega} d^3{\vec x} \hspace{0.06in} \left(\rho_0\ +
:\pi^*({\vec x}) \pi({\vec x}) + \partial_i \phi^*({\vec x}) \partial_i
\phi({\vec x}):\right),
\end{equation}
where $\pi^*({\vec x})$ is the conjugate momentum of the scalar field:

\begin{equation}
\label{momentum}
\pi^{*}({\vec x}) = \int \frac{d^3{\vec k}}{(2\pi)^3}
{\sqrt{|{\vec k}|}} P^*_{\vec k}
e^{-i{\vec k}\cdot{\vec x}}
\end{equation}
and $\rho_0 \neq 0$ is the average energy density. Only for the sake of 
elegance we have separated the average energy density as a free parameter 
in (\ref{hamiltonian}) and canceled out the expectation value 
of the second term, enclosed between signs $:--:$ of normal ordering.

The conjugate operators $Q_{{\vec k}_1}$ and $P^*_{{\vec k}_2}$ obey 
canonical commutation relations $[Q_{{\vec k}_1}, P^*_{{\vec k}_2}] = i\ 
(2\pi)^3 \delta^3({\vec k}_1 - {\vec k}_2)$.
Introducing the Fourier expansions (\ref{field}) and (\ref{momentum}) in 
(\ref{hamiltonian}) we can describe the dynamics of the scalar field as a 
collection ${\cal H} = {\cal H}_0\ +
:\int_{k_{IR}} \frac{d^3{\vec k}}{(2\pi)^3} \hspace{0.05in} |{\vec k}| 
\left[P^*_{\vec k} P_{\vec k} + Q^*_{\vec k} Q_{\vec k}\right]:$ of 
free harmonic oscillators. The zero mode is $ {\cal H}_0 \equiv 
\left(\int_{\Omega} d^3{\vec x}\ \rho_0\right)$. 

If the expansion of the space-time background preceding the time of 
matter radiation equality is perfectly adiabatic, the scalar quantum field 
reaches that instant in its fundamental state $|0 \rangle$. In other 
words, we assume here that there is not cosmological production of real 
$\phi$-particles. The ground state of the scalar field is 
the tensorial product of the vacua of the decoupled harmonic modes. 
Fourier modes do fluctuate quantum mechanically in their 
ground state $P_{\vec k}|0 \rangle = i\ Q_{\vec k}|0 \rangle$ with normal 
distribution of covariance $\langle Q^*_{{\vec k}_1} Q_{{\vec k}_2} 
\rangle= (2\pi)^3 \delta^3({\vec k}_1 - {\vec k}_2)$ and can generate 
gaussian fluctuations in the macroscopic restricted observables ${\cal 
H}_V = \int_V d^3{\vec x} \hspace{0.07in} {\cal H}({\vec x})$ that
describe the energy (mass) contained in macroscopic spatial sub-volumes 
$V$,

\begin{eqnarray}
\nonumber
\label{restricted}
{\cal H}_V = 
\rho_0 V + :\int_V d^3{\vec x} \int_{k_{IR}} \frac{d^3{\vec 
k}_1}{(2\pi)^3} \frac{d^3{\vec k}_2}{(2\pi)^3}
\left(\sqrt{|{\vec k}_1||{\vec k}_2|} P^*_{{\vec k}_1}
P_{{\vec k}_2} + \frac{{\vec k}_1\cdot{\vec k}_2}{\sqrt{|{\vec
k}_1||{\vec k}_2|}} Q^*_{{\vec k}_1} Q_{{\vec k}_2}\right)
e^{-i({\vec k}_1-{\vec k}_2)\cdot{\vec x}}:,
\end{eqnarray}
because the ground state $|0 \rangle$ of the quantum field is not an 
eigenstate of these macroscopic bulk operators.

The typical size of the macroscopic fluctuations is estimated by the 
variance $(\Delta E_V)^2 \equiv \langle 0|{\cal H}_V^2|0 \rangle - \langle 
0|{\cal H}_V|0 \rangle^2$. A formal expansion of this expression was 
obtained in \cite{Brustein:2000hh} and rederived in \cite{collar},

\begin{equation}
\label{variancefield}
(\Delta E_V)^2 = \frac{1}{4(2\pi)^6} 
\int^{k_{UV}}_{k_{IR}} d^3{\vec k}_1
d^3{\vec k}_2 \hspace{0.015in} |F_V({\vec k}_1-{\vec
k}_2)|^2 |{\vec k}_1| |{\vec k}_2|
\left(cos(\theta) - 1\right)^2. 
\end{equation}
where $k_{UV} \simgt H(t_{eq})$ is some cutoff that regularizes the theory
in the ultraviolet and $cos(\theta)\equiv \frac{{\vec k}_1\cdot{\vec k}_2}
{|{\vec k}_1| |{\vec k}_2|}$ is the cosine of the angle opened between 
the two vectors ${\vec k}_1$ and ${\vec k}_2$. 
The need for a regularization ultraviolet cutoff to render the variance 
(\ref{variancefield}) finite was already 
encountered in our discussion in Section III, case I. In next section, 
we will discuss this regularization procedure in the context of a 
renormalization program of the physical parameter that measures the size 
of the mass anisotropies. 

Comparing (\ref{variancefield}) with the general expression
(\ref{variance})

\begin{equation}
\label{variancefield2}
(\Delta E_V)^2 = \rho^2_0 \int^{2k_{UV}}_0 \frac{d^3{\vec
\zeta}}{(2\pi)^3} \hspace{0.05in} {\cal P}(|{\vec \zeta}|)\
|F_V({\vec \zeta})|^2,
\end{equation}
where the geometric factor $F_V({\vec \zeta})$ is given in expression
(\ref{geometry}), we can easily obtain the power spectrum of vacuum 
fluctuations:

\begin{equation}
\label{powerspectrumQFT}
{\cal P}(|{\vec \zeta}|) = \frac{1}{\rho^2_0} \frac{1}{32 (2\pi)^3}
\int_{d({\vec \zeta})} d^3{\vec \mu}\hspace{0.015in} 
|{\vec k}_1| |{\vec k}_2| \left(cos(\theta) - 1\right)^2,
\end{equation}
in terms of the new variables ${\vec k}_1=\frac{1}{2}\left({\vec
\mu}+{\vec \zeta}\right)$ and ${\vec k}_2=\frac{1}{2}\left({\vec
\mu}-{\vec \zeta}\right)$. The domain of integration in momentum space is
defined by the condition $d({\vec \zeta}) \equiv \{{\vec \mu} \in R^3:
k_{IR} \le \frac{1}{2}|{\vec \mu} \pm {\vec \zeta}| \le k_{UV} \}$.
This domain is the intersection region of two similar annulus with inner
radius $2 k_{IR}$ and outer radius $2 k_{UV}$ and centered, respectiveley,
in $\pm {\vec \zeta}$.
\\

It is easy to check that at $\zeta=0$, that is ${\vec k}_1 = {\vec k}_2$, 
the power spectrum vanishes

\begin{equation}
\label{condition}
{\cal P}(\zeta=0) = lim_{\zeta \rightarrow 0}\ P(\zeta) = 0.
\end{equation}
This result is a direct consequence of the fact that the quantum field is
in its ground state, which is an eigenstate of the hamiltonian
(\ref{hamiltonian}) and, therefore, the total energy $E(\Omega)$ in the
whole system $\Omega$ does not fluctuate quantum mechanically, $(\Delta
E_{\Omega})^2 = 0$. This condition simply constrains the fluctuations 
of the density field to preserve the total energy (mass) in the system 
and, according to (\ref{implication}), it is sufficient to {\it force} 
the energy (mass) anisotropies (\ref{variancefield}) to be scale 
invariant (\ref{scaleinvariance}) over cosmologically large covariant
sub-volumes \cite{Oaknin:2003dc}, $(\Delta E_V)^2 \propto S$.

In fact, the scale invariance of anisotropies (\ref{variancefield}) over 
cosmologically large comoving volumes was proved numerically in 
\cite{Brustein:2000hh}. Additional numerical examples that beatufilly 
prove the claim were presented in \cite{collar}. In a first example, the 
variance of energy fluctuations of a free scalar field in $2+1$ Minkowski 
space-time is calculated in the spatial region within a wrinkled surface. 
The interesting aspect of this example comes from the fact that the total 
volume contained within the surface does not change when the boundary 
surface wrinkles, although the area of the boundary grows monotonically. 
The results presented there show that the variance of energy fluctuations 
does grow linearly with the growing boundary surface. In a second example, 
the variance of energy fluctuations is calculated within an annulus of 
inner radius $r_1$ and outer radius $r_2$. When the radius $r_1$ gets 
larger, but still smaller than $r_2$, the total volume of the annulus 
decreases. But the area of the boundary surface grows, and so also does 
the variance of energy fluctuations. 

When this mechanism of local rearrangement is discussed in the literature 
the attention is focused on the quadratic coupling in the hamiltonian 
density (\ref{hamiltonian}) of Fourier modes of the fields $\phi({\vec 
x})$ or $\pi^*({\vec x})$ with close covariant momenta ${\vec k}_1 \simeq 
{\vec k}_2$, which produces fluctuations of the density field ${\cal 
H}({\vec x})$ with covariant wavelength $\lambda = 2\pi/|{\vec \zeta}| =
2\pi/|{\vec k}_1-{\vec k}_2|$ that can be much longer than the horizon,
even if both covariant wavelengths $2\pi/|{\vec k}_1|$ and $2\pi/|{\vec
k}_2|$ are much shorter than the horizon. This is a very simple property
of the coupling of Fourier modes that led to Y.B.~Zeldovich, and many 
others after him, to notice many years ago \cite{YaZeldovich} that local 
fluctuations of the scalar field (\ref{field}) with comoving wavelength 
shorter than the horizon can produce a non zero power spectrum 
(\ref{powerspectrumQFT})

\begin{equation}
{\cal P}(|{\vec \zeta}|) \sim \frac{1}{\rho^2_0} \frac{1}{2
(2\pi)^2} k_{UV} |{\vec \zeta}|^4
\end{equation}
over cosmologically short comoving momenta $\zeta \simlt H(t_{eq})$. 

The interest in the literature on this aspect of the mechanism is 
motivated, nevertheless, by the erroneous assumption 
(\ref{preliminaries}), which wrongly asserts that mass (energy) 
anisotropies (\ref{variancefield2}) over cosmologically large comoving 
volumes of size $L \gg H^{-1}(t_{eq})$ are generated by the contribution 
$d\zeta\ \zeta^2\ {\cal P}(\zeta) |F_V(\zeta)|^2$ at $\zeta \sim L^{-1} 
\ll H(t_{eq})$. This erroneous assumption (\ref{preliminaries}) has 
misled the analysis of the mechanism of local rearrangement of matter as 
the origin of cosmological anisotropies since the early years when it was 
first studied.
As we explained in Section III, case I), the contribution at $\zeta \sim 
L^{-1}$ to (\ref{variancefield2}) is negligible compared to the 
ultraviolet contribution, $d\zeta\ \zeta^2\ {\cal P}(\zeta) 
|F_V(\zeta)|^2$ at $\zeta \sim k_{UV}$, in statistical systems globally 
constrained by the condition $(\Delta M_{\Omega})^2 = 0$. This can be 
checked directly on (\ref{variancefield}) by noticing that the largest 
contribution comes from $|{\vec k}_{1,2}| \sim k_{UV}$ when they are 
oppositely oriented, ${\vec k}_1 \sim -{\vec k}_2$, so that $cos \theta = 
-1$, which contribute to the power spectrum ${\cal P}(\zeta \gg 1/L)$ in 
the ultraviolet range, and not by the aligned modes ${\vec k}_1 \sim {\vec 
k}_2$, which form an angle $cos \theta \sim +1$ and contribute to
${\cal P}(\zeta \sim 1/L \ll H(t_{eq}))$. When the dominant ultraviolet 
contribution is taken into account (\ref{tres}) the resulting 
anisotropies \cite{collar} over cosmologically large volumes are proven 
numerically to be scale invariant.

The irrelevance of the quadratic dependence of the density field 
(\ref{hamiltonian}) in the fundamental fields (\ref{field}) and 
(\ref{momentum}) for generating scale invariant anisotropies over 
cosmologically large volumes out of fluctuations in Fourier modes of the 
latter with very short comoving wavelength is manifest in the next example.

Let us consider the same theoretical framework laid in (\ref{action}),
(\ref{hamiltonian}), but assume now that the matter density field is 
proportional to (\ref{momentum}) the conjugate momentum field. 
Obviously, the density field (\ref{momentum}) is still homogeneous and 
isotropic on average:

\begin{equation}
\label{averagehomiso}
\langle 0|\pi(\vec x)|0 \rangle = 0.
\end{equation}
But, as we know, this constrain cannot prevent the appearance of 
random quantum anisotropies in the spatial energy (mass) density 
distribution, because the ground state $|0 \rangle$ is 
not an eigenstate of the restricted operators

\begin{equation}
\label{PIm}
M_V \propto \int_V d^3{\vec x}\ \pi({\vec x}) + {\it h.c.}.
\end{equation}
if the volume of intergration $V$ is not the whole 3D space $\Omega$.
 
We find the variance of energy (mass) anisotropies in finite comoving 
volumes of cosmological size to be

\begin{equation}
\label{vardefp}
(\Delta M_V)^2 \propto \int_V \int_V d^3{\vec x}\ d^3{\vec y} 
\langle 0|\pi^*(\vec x) \pi(\vec y)|0 \rangle = \int^{k_{UV}}_{k_{IR}} 
\frac{d^3{\vec k}}{(2\pi)^3}\ k\ |F_V({\vec k})|^2,
\end{equation}
given that $\langle 0|\pi^*(\vec x) \pi(\vec y)|0 \rangle = \int 
\frac{d^3{\vec k}}{(2\pi)^3} k\ e^{i {\vec k}\cdot({\vec x} - {\vec y})}$.
Now, the density field depends linearly in the fundamental fields, 
which are allowed only to randomly fluctuate in modes with comoving 
wavelength shorter than the horizon. We read from the expression above 
the power spectrum of the fluctuations in this example, ${\cal P}(k) 
\sim k$ for $k_{IR} \le k \le k_{UV}$ and is zero otherwise.

The ultraviolet cutoff $k_{UV}$ is necessary to render expression 
(\ref{vardefp}) finite. Once we introduce the cutoff we immediately 
confirm, following previous discusssion in Section III, case II): 

\begin{equation}
\label{vardef3}
(\Delta M_V)^2 \sim L^2
\end{equation}
quantum anisotropies of the density field (\ref{PIm}) are scale 
invariant.
This result was foreseeable on the ground $(\Delta M_{\Omega})^2 = 0$, 
simply expressing that also in this second example total energy (mass) is 
conserved and not allowed to fluctuate quantum  mechanically. The need 
for an ultraviolet cutoff in (\ref{vardefp}) proves that these 
anisotropies are also mainly generated by the contribution from the modes 
with the shortest comoving wavelength. 

The two examples discussed in this section basically consist of
random fluctuations that can displace the carriers of mass (energy), {\it 
i.e.} fluctuations of the fundamental fields (\ref{field}) and 
(\ref{momentum}), only over causally connected distances. They both generate 
scale invariant mass (energy) anisotropies over cosmologically large 
comoving volumes. On the other hand, the two examples are defined by 
completely different density power spectra. The power spectrum associated 
to (\ref{variancefield}) density fluctuations of the quadratic operator 
${\cal H}_V$ is ${\cal P}(k) \sim k^4$ over the whole range of 
cosmologically short momenta $k \simlt k_{UV}$, with $k_{UV} \simgt 
H(t_{eq})$, while the power spectrum associated to (\ref{vardefp}) 
fluctuations of the momentum density operator $M_V$ in the second example 
is ${\cal P}(k) \sim k$ in the range of momenta within the horizon 
$H(t_{eq}) \simlt k \simlt k_{UV}$. We already found this degeneracy in 
the power spectrum of scale invariant anisotropies in our previous 
analysis in Section III, case I).

We close this section with a short comment on the cutoff procedures we 
have introduced above in this section. First, we realize that for the 
issues we have been discussing here the infrared cutoff $k_{IR}$ is 
irrelevant in both examples, as the contribution of the long 
wavelength modes to the total variance (\ref{variance}) is sub-dominant. 
The dominant contribution comes in both examples from the short  
wavelength modes of the fundamental fields. In the first example, 
(\ref{variancefield}), the 
dominant contribution is obtained from the quadratic coupling of 
oppositely oriented ultraviolet modes $|k_{1,2}| \sim k_{UV}$, $cos\theta 
\simeq -1$. In the second example, the largest contribution to (\ref{vardefp}) 
comes also from the ultraviolet modes of the momentum field. Both
contributions are divergent and needs to be regularized with ultraviolet 
cutoffs. Now it is important to notice from (\ref{integral3}) that the 
ultraviolet regularization procedure only affects the absolute size of 
the fluctuations, $(\Delta M_V)^2 \sim \rho^2_0\ \mu S$, but not the 
linear dependence on the area $S$ of the surface that bounds the 
sub-volume. It is natural then to understand this procedure as an 
intermediate regularization step in the renormalization programme of 
the physical parameter $\mu = \frac{{\cal A'} k^n_{UV}}{k^{n+3}_c}$.
This comment introduces us to the issues addressed in next Section.

\section{Scale invariant density anisotropies in quantum field theories.}

The scalar quantum field theory (\ref{action}) of last section is 
obvioulsy only a toy-model designed to confront some widespread, but 
wrong, beliefs about the dynamics of primordial structure formation. 
The model has also been a satisfactory bench to discuss some 
important concepts that shall be very relevant in the development of
a complete quantum field theory of primordial cosmological density 
perturbations.

Within the setup of a QFT the energy (mass) density 
field at the time of equality (\ref{density}) is described by a density 
operator

\begin{equation}
\label{densityQ}
{\it h}({\vec x}; t_{eq}) = {\it h}_0 + \langle {\it h}_0 \rangle 
\int \frac{d^3{\vec k}}{(2\pi)^3}\ {\delta}_{\vec k}\
e^{-i{\vec k}\cdot{\vec x}},
\end{equation}
analogous to the density operator in (\ref{hamiltonian}),
acting on the quantum state of the system $| 0 \rangle$. This state 
is required to be invariant under spatial translations and rotations.
Homogeneity and isotropy demand $\langle 0|{\it h}({\vec x})|0 \rangle 
= \langle 0|{\it h}_0|0 \rangle$, which implies

\begin{equation}
\label{homogeneityQ}
\langle 0| {\delta}_{\vec k}|0 \rangle = 0,
\end{equation}
and, therefore, the average value of $M(V) \equiv \int_V d^3{\vec 
x}\ {\it h}({\vec x})$, the energy (mass) contained 
within a spatial macroscopic sub-volume $V$, is proportional to the volume 
of the region, $\langle 0|M(V)|0 \rangle = V \times \langle 0|{\it 
h}_0|0 \rangle$. The magnitude $M(V)$ is said to be an extensive 
properties of the system. We assume here, as we did in Section III, that  
$\langle {\it h}_0 \rangle \equiv \langle 0|{\it h}_0|0 \rangle \neq 0$. 

Condition (\ref{homogeneityQ}) does not prevent, nevertheless, the 
appearance of random spatial energy (mass) density anisotropies due to 
quantum fluctuations, with variance 

\begin{equation}
\label{varianceQ}
(\Delta E_V)^2 \equiv \langle 0|\left[M(V) - \langle M(V) 
\rangle\right]^2|0 \rangle = {\langle {\it h}_0 \rangle}^2 \int 
\frac{d^3{\vec k}}{(2\pi)^3}\ {\cal P}(k) |F_V({\vec k})|^2,
\end{equation}
where the power spectrum ${\cal P}({\vec k})$ is defined, as usual, by

\begin{equation}
\label{powerQ}
\langle 0|{\delta}^*_{{\vec k}_1}\ {\delta}_{{\vec k}_2}|0 \rangle =
(2\pi)^3 {\cal P}({\vec k}_1) \delta^3({\vec k}_1 - {\vec k}_2). 
\end{equation}
In a wider context, (\ref{densityQ}) can be the density field operator of 
any global Noether charge

\begin{equation}
\label{hamiltonianQ}
{\cal H} = \int_{\Omega} {\it h}({\vec x}) = \int_{\Omega} {\it h}_0.
\end{equation}
The physical observable $M(V)$ describes then the net value of that charge 
within a finite macroscopic comoving sub-volume of cosmological size. 
Condition (\ref{homogeneityQ}) implies that charge ${\cal H}$ is, on 
average, homogeneously and isotropically distributed in space, while
(\ref{varianceQ}) reminds us that quantum fluctuations can, nevertheless, 
produce random anisotropies in its spatial distribution. 

According to the central limit theorem, for a complete 
statistical description of random anisotropies in the spatial 
distribution of the extensive charge is enough to provide the value of 
their variance (\ref{varianceQ}) over comoving volumes of arbitrary 
cosmological size. The calculation of the variance of random spatial 
anisotropies in the distribution of a Noether charge is a well-posed 
physical question that should get an unambiguous, finite answer in any 
properly formulated theoretical framework. Of course, the theoretical 
calculation must produce the clear result $(\Delta E_V)^2 = 0$ over any 
macroscopically large comoving volume if and only if the system does not 
generate random anisotropies. 

Hence, it is intriguing to notice that within the setup that we are 
considering we are allowed to modify  the definition of 
the charge density field operator $(\ref{densityQ})$, for example by 
adding a total three-divergence, without changing the definition of the 
global charge (\ref{hamiltonianQ}). Such 'symmetry' transformations of the 
charge density field (\ref{densityQ}) must 
preserve the definition of the zero-mode ${\it h}_0$ and they must also 
respect the expectation values (\ref{homogeneityQ})  of the non-zero 
Fourier modes $\delta_{\vec k}$, ${\vec k} \neq 0$ in order to preserve 
the average homogeneity and isotropy of the system, but they can alter the 
power spectrum ${\cal P}(k)$ at $k \neq 0$, defined in (\ref{powerQ}).
Therefore, they can alter the variance of quantum mechanically generated 
spatial charge anisotropies in finite macroscopic sub-volumes, defined in 
(\ref{varianceQ}). This degeneracy implies, in principle, that these two 
concepts (\ref{powerQ}) and (\ref{varianceQ}) cannot be uniquely defined 
in the considered theoretical framework and raises questions on the 
capability of the setup to properly describe quantum 
mechanically generated spatial anisotropies. 
In the paragraphs below we investigate and clarify this point.

We need first to identify the invariant features of the power spectrum 
$(\ref{powerQ})$ and variance (\ref{varianceQ}) under the 'symmetry' 
transformations of the density field (\ref{densityQ}): as the zero mode 
${\it h}_0$ is not modified by the permitted transformations of the 
density field operator, its statistical momenta $\langle 0|\left({\it 
h}_0 - \langle{\it h}_0\rangle\right)^p|0 \rangle$, $p=1,2,3,...$, are 
invariant features. The first moment $p = 1$ is zero, by definition.
The second and higher momenta, on the other hand, are zero only if the 
quantum state $|0 \rangle$ is an eigenstate of the global charge 
(\ref{hamiltonianQ}). In particular, this implies that energy (mass) is 
globally conserved, 

\begin{equation}
\label{vacuum}
(\Delta E_{\Omega})^2 = 0,
\end{equation}
by quantum random fluctuations of the density field (\ref{densityQ}).

This condition (\ref{vacuum}) is naturally satisfied if $|0 \rangle$ is 
the ground state of the system and (\ref{densityQ}) is a conserved 
Noether charge. According to our result (\ref{implication}) in section IV 
such constrain implies that quantum mechanically 
generated anisotropies (\ref{varianceQ}) in the spatial distribution of 
charge ${\cal H}$ must be scale invariant (\ref{scaleinvariance}) over 
comoving volumes of cosmological size. Thus, the linear dependence of the 
variance of primordial cosmological energy (mass) anisotropies on the 
area of the surface that bounds the considered comoving volume, 

\begin{equation}
\label{renormalization}
Ln\ (\Delta E_{V_1})^2 = Ln\ (\Delta E_{V_2})^2 + 
Ln\left(\frac{S_1}{S_2}\right), 
\end{equation}
where $V_1$ and $V_2$ are arbitrary macroscopic sub-volumes and $S_1$, 
$S_2$ are, respectively, the areas of the surfaces that bounds each of 
them, is an invariant feature under 'symmetry' transformations of the 
density field operator in the ground state of the system 
\cite{Oaknin:2003dc}. 
This is nothing but an example of the basic feature that we already 
noticed in our analysis of Section III, case I), that power 
spectra with different spectral index $n > 1$ cannot be distinguished 
through the energy (mass) density anisotropies they produce over 
macroscopically large comoving volumes: 'symmetry' tranformations of the 
density field (\ref{densityQ}) simply modify the power spectrum 
(\ref{powerQ}) of spatial anisotropies ${\cal P}(k) \rightarrow {\cal 
P}(k) + \Delta{\cal P}(k)$ by a regular term $\Delta{\cal P}(k) = 
o(k)$, which does not alter the macroscopic scale invariance 
(\ref{scaleinvariance}) of the anisotropies.

The transformations of the density field (\ref{densityQ}) can modify, on 
the other hand the size of the spatial charge anisotropies 
over macroscopically large spatial sub-volumes, parameterized 
by the factor of proportionality $\mu$,

\begin{equation}
\label{parameterRENORM}
(\Delta E_V)^2 \simeq \langle {\it h}_0 \rangle^2\ \mu S.
\end{equation}
This parameter $\mu$ is not an invariant feature under 'symmetry' 
transformations in the definition of the charge density and we should 
conclude that it describes a physical property 
that is not properly defined in this setup. We explain below the origin 
of this problem and argue, using usual statistical/QFT renormalization 
concepts, that this parameter $\mu$ must be considered as an additional 
free parameter of the theory, like masses or coupling constants, which 
fixes the size of quantum mechanical random spatial charge anisotropies 
in the ground state of the system. 

In Section III, case I), we learned that scale invariant energy (mass) 
anisotropies over macroscopically large volumes are dominated by the 
contribution of the ultraviolet modes to the integral expression 
(\ref{integral}). To render this expression finite we introduced by 
hand an ultraviolet cutoff, which fixes the size of the mass (energy) 
anisotropies in cosmologically large comoving volumes. We can now 
understand this step in Section III as a regularization procedure of the 
infinite parameter $\mu$. The 'symmetry' transformations 
of the density field (\ref{densityQ}) can be reabsorbed within this 
infinite parameter. Or, in a different point of view, the infinite 
parameter $\mu$ can be regularized by adding an appropriate counter-term 
in the density field. This procedure do not modify the finite 
relationship (\ref{renormalization}), which is a physical prediction of 
the setup. In conclusion, scale invariance of vacuum anisotropies, 
expressed through the finite relationship (\ref{renormalization}), must 
be respected through the renormalization program. The price we pay is 
the normal price in QFT/statistical mechanics: we cannot predict from 
first principles within this setup the value of the 
'dressed' parameter $\mu$, in the same way that we cannot predict the 
physical value of $\alpha_{em}$ in QED. 

Only after fixing, normally 
through measurement, the variance $(\Delta E_{V_1})^2 = \Delta^2$, 
of random anisotropies over a macroscopic volume of reference $V_1$, 
we have a meaningful physical prediction

\begin{equation}
(\Delta E_{V_2})^2 = \frac{S_2}{S_1} \Delta^2,
\end{equation}
for the variance of mass (energy) anisotropies over any other 
comoving volume $V_2$ of cosmological size.
This relationship is equivalent to the renormalization group equation that 
predicts physical values for the running of the coupling constant 
$\alpha_{em}(p^2)$ at arbitrary energy scales only after an initial value 
$\alpha_{em}(p_0^2)$ at a certain renormalization energy scale has been 
fixed by experiment.

Please notice that the incapability of the QFT setup to predict the 
'dressed' value of the physical parameter $\mu$ due to common ultraviolet 
divergences and the impossibility to single out in the setup the local 
definition of the energy (mass) density operator, reaches also the 
widely accepted {\it predictions} of the inflationary scenario. We can 
freely add a three-divergence to the hamiltonian density, which will {\it 
dress} the scale invariant anisotropies {\it predicted} by inflation. 
As we have said, the additional terms respect the scale 
invariance of the anisotropies (\ref{renormalization}), but make the 
parameter $\mu$ in (\ref{parameterRENORM}) formally infinite.
Let us say it in different words: under mistaken assumption 
(\ref{berlusco}), upon which the calculations of the 
cosmological anisotropies in inflationary cosmology are carried out, there 
are no divergences in the theoretical calculation of the anisotropies, but 
just because the infinites appear only after noticing the degeneracy 
(\ref{tres}) in the power spectrum of scale invariant mass (energy) 
cosmological anisotropies. 
Ordinary predictions of inflationary cosmology of primordial structure 
does not suffer from the problem of infinite expressions, only because in 
these calculations the infinities are thrown away, using approximation 
(\ref{berlusco}), without justification. A correct, not based on mistaken 
assumption (\ref{berlusco}), complete calculation of inflationary 
predictions will also lead to finite relationship 
(\ref{renormalization}), but only after a regularization procedure of the 
infinite expression for the parameter $\mu$. 

It is very interesting to notice that the infinite contribution from 
ultraviolet modes of the density field to the physical parameter $\mu$ in 
(\ref{parameterRENORM}) is, according to (\ref{implication}),
 directly associated to the fact that total mass (energy) is globally 
conserved $(\Delta E_{\Omega})^2 = 0$. Scale invariant anisotropies of 
extensive magnitudes defined in the bulk of an spatial volume $V$ are 
associated to the surface that bounds the sub-volume. If these 
anisotropies can be {\it holographically} described by a theory defined 
on the surface, as suggested in \cite{Brustein:2003pt},\cite{collar}, 
we interestingly find a connection between the infinite, 
renormalizable parameter $\mu$ of the {\it holographic} theory defined on 
the closed surface, and the globally conserved extensive magnitude 
defined in the interior and exterior volumes to that surface. 
In quantum Hall effect of strongly correlated electrons confined in 2D 
spatial dimensions by a perpendicular magnetic field in condensed matter 
physics it is known that is possible to obtain an {\it holographic} 
1D description of the edge excitations over the boundary border 
\cite{Halperin}.

The problem of determining the size $\mu$ of scale invariant energy (mass)
anisotropies (\ref{parameterRENORM}) is not very different from the 
problem of determining the cosmological constant in the context of 
relativistic QFTs. The average value $\langle M(V) \rangle =  V 
\times \langle {\it h}_0 \rangle$ is, in general, ultraviolet divergent. 
We cannot predict within the QFT setup the value of $\Lambda \equiv 
\langle {\it h}_0 \rangle$, but once the theory 
is regularized we find that $\langle E_{V_1} \rangle = 
\frac{V_1}{V_2} \langle E_{V_2} \rangle$, the energy (mass) 
contained within a macroscopic sub-volume is an extensive magnitude.
This is a finite relationship that must be respected whatever the unknown
mechanism that renormalizes the value of $\Lambda$ could be.
In the same way we cannot predict the physical value of the 
parameter $\mu$, but we can predict the scale invariance 
(\ref{renormalization}) of the variance of spatial charge anisotropies. 
Both of these problems become physically meaningful only when 
the energy (mass) density field gets coupled to gravity and, therefore, 
they can only get a final answer in the more fundamental setup that 
includes the quantum description of both matter components and dynamics 
of the space-time metric.

\section{Scale invariant anisotropies in the gauge invariant formalism 
of linearized density perturbations.}

Spatial anisotropies in the energy-momentum density field necessarily
perturbate the homogeneous and isotropic FRW expanding space-time 
background, which in turn can feed back the dynamics of the density 
anisotropies. A complete analysis of the locally Lorentz invariant 
equations of general relativity that describe the coupled dynamics of 
density and metric perturbations is technically very demanding even at 
the classical level because it must carefully take into account 
unphysical gauge degrees of freedom. Fortunately, a first aproximation 
of the coupled dynamics at linear order in the perturbations has been 
formulated in a very simple gauge invariant framework. A compiling 
detailed report of this approach can be found in 
\cite{Mukhanov:1990me}. The linear approach is justified in 
the literature on the argument that cosmological primordial mass 
(energy) perturbations are tiny, $\delta \rho/\rho \simlt 10^{-5}$. 
In the previous section we have found, nevertheless, that in the 
standard statistical setup the physical parameter $\mu$, which measures 
the size of the fluctuations, $\delta \rho \propto \sqrt{\mu\ S}$, 
is infinite and needs to be renormalized. This brings us to face another 
usual procedure in perturbative QFT: the perturbative expansion must be 
carried out in powers of the {\it dressed} physical value of the
parameter $\delta \rho/\rho$, which is really tiny, and not in 
terms of the infinite, or very large, {\it undressed} parameter.  
We will not further discuss here how to carry a renormalization 
programme. In this section we only explore 
within the gauge invariant framework at linear order the concepts 
introduced in the previous sections. This section VII, like Section VI, 
report only first results of research in progress.

First, let us very briefly review the basics of the linearized 
gauge invariant formalism, which start from the usual Einstein-Hilbert
action for the coupled dynamics of the space-time metric and 
energy-momentum matter tensors. Both tensor field are then decomposed 
as a sum of two terms: a first homogeneous and isotropic FRW background 
term plus a second perturbative term describing the anisotropies.
Using the equations of motion the action is rewritten in linear 
aproximation keeping up to second order in the perturbative terms.
After some straightforward, but lengthty, calculations the general action 
of coupled density and metric perturbations during the radiation 
dominated epoch of a FRW background is written in comoving 
coordinates in terms of a single gauge invariant scalar field 
potential $v({\vec x})$, whose dynamics is described by the hackneyed 
hamiltonian of a free scalar field in flat Minkowski space-time 
\cite{Mukhanov:1990me}

\begin{equation}
\label{lineardensity}
{\cal H} = \int_{\Omega} {\cal H}({\vec x}) = \int_{\Omega} d^3{\vec x} 
\left(\pi({\vec x})\pi({\vec x}) + \frac{1}{3}
{\vec \nabla}v({\vec x}){\vec \nabla}v({\vec x})\right). 
\end{equation}  
Field operator $v({\vec x})$ and its conjugate momentum $\pi({\vec y})$ 
obey canonical commutation relations $[v({\vec x}),\pi({\vec y})] = 
i\ \delta^3({\vec x}-{\vec y})$. 

During the radiation dominated stage of the standard cosmology there 
is not cosmological particle production and, therefore, the gauge 
invariant scalar field remains in its ground state $|0 \rangle$ when 
the temperature of the universe cools down to the onset of matter 
radiation equality. 

Furthermore, every physical observable can be expressed in this 
formalism in terms of this pair of conjugate fields, thus eliminating 
the problem of spurious gauge degrees of freedom. In particular, the 
Fourier modes of the energy (mass) density field operator 
(\ref{densityQ}) with comoving wavelength shorter that the horizon 
are proportional in this formalism to the Fourier modes of the momentum 
density operator:

\begin{equation}
\label{momentumDF}
\frac{1}{\langle {\it h}_0 \rangle}({\it h}({\vec x}) - {\it h}_0) 
\simeq \frac{1}{T_{eq}^2}\ \left(\pi({\vec x}) + {\it h.c.}\right) + 
\partial_i J^i({\vec x}),
\end{equation}
where $T_{eq}$ is the cosmic temperature at the time of equality and 
$\partial_i J^i({\vec x})$ is an undefined arbitrary three divergence 
according to our discussion in the previous section.
Formally, this density field is the second of the two examples we  
discussed in the toy-model of Section V. We found there that in the 
ground state of the system $|0 \rangle$ its spatial anisotropies 
are scale invariant (\ref{vardefp}). 

We consider important to remark that the density field operator in the 
linearized gauge invariant formalism is proportional to the momentum 
density (\ref{momentumDF}) only over causally connected distances. Over 
super-horizon distances they do not necessarily coincide and the right 
hand side of (\ref{momentumDF}) gets correction terms. This is not really 
important to us, as we know that the largest contribution to scale 
invariant cosmological anisotropies, globally constrained to preserve the 
total energy (mass) of the system, comes from fluctuations in 
under-horizon modes of the density field (\ref{integral3}). These modes 
are described by (\ref{momentumDF}).

The unsolved situation regarding the theoretical prediction of the 
parameter $\mu$ in (\ref{parameterRENORM}) has not improved in this 
formalism of linearized density anisotropies coupled to metric 
perturbations. A direct estimation of the variance of energy (mass) 
anisotropies from (\ref{momentumDF}) produces

\begin{equation}
(\Delta E_V)^2 \simeq \langle {\it h}_0 \rangle^2 \frac{1}{T^4_{eq}} S.
\end{equation}
Over a sphere of comoving radius $L$ we then obtain

\begin{equation}
\left(\frac{\delta \rho}{\rho}\right)_V^2 = \frac{(\Delta E_V)^2}{\langle 
{\it h}_0 \rangle^2 V^2} \simeq \frac{9}{4\pi} \frac{1}{(T_{eq} L)^4}.
\end{equation} 
A rapid estimation for $L \sim 10^3 H^{-1}_{eq} \sim 10^3 
\frac{T_{eq}}{M^2_P}$ gives for the {\it undressed} dimensionless 
parameter $\left(\frac{\delta \rho}{\rho}\right)_V
\sim \sqrt{\frac{9}{4\pi}} 10^{-6} \left(\frac{M_P}{T_{eq}}\right)^4 
\sim 10^{100}$. Obviously, this is much larger than the observed physical
{\it dressed} value $\left(\frac{\delta \rho}{\rho}\right)_V \simlt 
10^{-5}$. The renormalization can be carried out through counter-terms 
introduced by the three-divergence in the definition of the density 
operator (\ref{momentumDF}).

For a deeper treatment of this problem is necessary to go beyond the 
linear order of the density and metric perturbations. 
We guess that the final quantum theory of gravity has a say in fixing 
the size of scale invariant primordial anisotropies. Dynamics of structure 
formation could thus offer a bench for phenomenologically testing that 
final theory at the low energies of matter-radiation
equality.

\section{Renormalization equations of physical density anisotropies.}

We wish to explore in this section how the renormalized ({\it
dressed}) mass (energy) anisotropies depend on the resolution  
scale at which the physical density field is probed and obtain, within the 
theoretical setup laid in section VI, the renormalization group equations 
that describe this dependence.

Let us say that $\rho_{\it phy}({\vec x}; {\it l}_0)$ is the physical
density field probed with arbitrary length resolution ${\it l}_0 =
M_0^{-1}$. When the system is probed at blunter length resolution 
${\it l}_1 = M_1^{-1} \simgt {\it l}_0$ (or, in energy scale $M_1 \simlt 
M_0$), the observable field

\begin{equation}
\label{smooth}
\rho_{\it phy}({\vec x}; {\it l}_1) = \int_{\Omega} d^3{\vec y}\ \rho_{\it
phy}({\vec y}; {\it l}_0)\ W({\vec y} - {\vec x}; {\it l}_1, {\it l}_0)
\end{equation}
looks further smoothed because its Fourier modes with wavelength $\lambda
\simlt {\it l}_1$ shorter than the size of the new probe are cut off
by an appropriate convolution kernel (window function) $W({\vec y} -
{\vec x}; {\it l}_1, {\it l}_0)$ extended around the point of observation
${\vec x}$ over a certain domain of typical size ${\it l}_1$ and
normalized to

\begin{equation}
\label{normalization}
\int_{\Omega} d^3{\vec y}\ W({\vec y} - {\vec x}; {\it l}_1, {\it l}_0) =
1.
\end{equation}
Expression (\ref{smooth}) simply means that measuring a local observable
at any point ${\vec x} \in \Omega$ with a probe of length ${\it l}_1$ does
in fact return an average weighted value of the field over a vicinity
of the size of the probe around the tested point. This is typically the
shortest scale that that probe can test.

Obviously, the physical density field (\ref{smooth}) resolved to
lengths of the order ${\it l}_1$ cannot depend on the arbitrary
shorter scale ${\it l}_0$ used in the definition above. This scale ${\it
l}_0$ plays here a role similar to the arbitrary renormalization scale in
renormalization group equations. The condition of independence of the
physical observable field (\ref{smooth}) resolved at length ${\it l}_1$ on
the renormalization scale ${\it l}_0$ fixes the functional dependence of
the convolution kernel

\begin{equation}
\label{runningwindow}
\int_{\Omega} d^3{\vec z}\ W({\vec y} - {\vec z}; {\it l}, {\it l}_0)\
W({\vec z} - {\vec x}; {\it l}_1, {\it l}) = W({\vec y} - {\vec x};
{\it l}_1, {\it l}_0).
\end{equation}
The simplest solution to this convolution equation is a Dirac delta
function $W({\vec x} - {\vec y}; {\it l}_a, {\it
l}_b) = \delta^3({\vec x} - {\vec y})$ for whatever two resolution length
scales ${\it l}_a, {\it l}_b$ we choose to compare, but it must be
understood that this choice is rather unphysical as it assumes that we can
solve the density field with infinitely sharp length resolution. A more
realistic solution to (\ref{runningwindow}) is a normalized gaussian
kernel:

\begin{equation}
\label{window}
W({\vec x} - {\vec y}; {\it l}_a, {\it l}_b) =
{\cal N} Exp[-\frac{|{\vec x} - {\vec y}|^2}{{\it l}_a^2 - {\it l}_b^2}]
= {\cal N} Exp[-\frac{M_b^2\ M_a^2}{M_b^2 - M_a^2} |{\vec x} -
{\vec y}|^2],
\end{equation}
where
\begin{equation}
{\cal N}^{-1} = \int_{\Omega} d^3{\vec y}\ Exp[-\frac{M_b^2\
M_a^2}{M_b^2 - M_a^2} |{\vec x} - {\vec y}|^2].
\end{equation}
Naturally, when $M_a \ll M_b$ the convolution kernel is $W({\vec x} -
{\vec y}; {\it l}_a, {\it l}_b) \simeq {\cal N} Exp[- M_a^2 |{\vec x} -
{\vec y}|^2]$. On the other hand, when we compare the observed
physical density fields at two very close resolutions scales ${\it l}_b
\simeq {\it l}_a$, the convolution kernel is roughly $W({\vec x} - {\vec
y}; {\it l}_a, {\it l}_b) \simeq \delta^3({\vec x} - {\vec y})$, as
it should be.

It is also useful to compare in momentum space the physical density fields
resolved at different length scales. Expanding $\rho_{\it phy}({\vec
x};{\it l}_1)$ and $\rho_{\it phy}({\vec x};{\it l}_0)$ in Fourier modes,
as in eq. (\ref{density}),

\begin{equation}
\label{dity0}
\rho_{\it phy}({\vec x};{\it l}_1) = \rho_0 + \rho_0 \int
\frac{d^3{\vec k}}{(2\pi)^3}\ {\delta}_{\vec k}[M_1]\
e^{-i{\vec k}\cdot{\vec x}},
\end{equation}
\begin{equation}
\label{dity1}
\rho_{\it phy}({\vec x};{\it l}_0) = \rho_0 + \rho_0 \int
\frac{d^3{\vec k}}{(2\pi)^3}\ {\delta}_{\vec k}[M_0]\
e^{-i{\vec k}\cdot{\vec x}}
\end{equation}
and then introducing the second expansion (\ref{dity1}) in eq.
(\ref{smooth})

\begin{eqnarray*}
\rho_{\it phy}({\vec x};{\it l}_1) =
\int_{\Omega} d^3{\vec y}\ \left(\rho_0 + \rho_0 \int \frac{d^3{\vec
k}}{(2\pi)^3}\ {\delta}_{\vec k}[M_0]\ e^{-i{\vec k}\cdot{\vec
y}}\right)\ W({\vec y} - {\vec x}; {\it l}_1, {\it l}_0) = \\
= \rho_0 + \rho_0 \int \frac{d^3{\vec k}}{(2\pi)^3}\ {\delta}_{\vec
k}[M_0]\ \left(\int_{\Omega} d^3{\vec y}\ e^{-i{\vec k}\cdot({\vec
y}-{\vec x})}\ W({\vec y} - {\vec x}; {\it l}_1, {\it l}_0)\right)
e^{-i{\vec k}\cdot{\vec x}}.
\end{eqnarray*}
Thus, we obtain by comparison with (\ref{dity0}) the relationship

\begin{equation}
\label{running}
{\delta}_{\vec k}[M_1] = {\widetilde W}_{\vec k}[M_1,M_0] {\delta}_{\vec
k}[M_0] = e^{-\frac{M_0^2-M_1^2}{4\ M_0^2\ M_1^2} k^2} {\delta}_{\vec
k}[M_0],
\end{equation}
where

\begin{equation}
\label{windowfunction}
{\widetilde W}_{\vec k}[M_1,M_0] = \int_{\Omega} d^3{\vec y}\ e^{-i{\vec
k}\cdot({\vec y}-{\vec x})}\ W({\vec y} - {\vec x}; {\it l}_1, {\it
l}_0) = e^{-\frac{M_0^2-M_1^2}{4\ M_0^2\ M_1^2} k^2},
\end{equation}
is the Fourier transform of the convolution kernel. We find
(\ref{running}) that, as a result of changing the scale at which we
resolve the density field, the stochastic modes $\delta_{\vec k}$ get
renormalized by the window function (\ref{windowfunction}). Therefore, the
power spectrum associated to these stochastic modes, defined by $(2\pi)^3 
{\cal P}({\vec k})\ \delta^3({\vec k}_1 - {\vec k}_2) = \langle 
\delta^*_{{\vec k}_1} \delta_{{\vec k}_2} \rangle$, gets renormalized as

\begin{equation}
\label{PSR}
{\cal P}_{\it phy}({\vec k}_1;M_1) = |{\widetilde W}_{{\vec
k}_1}[M_1,M_0]|^2\ {\cal P}_{\it phy}({\vec k}_1;M_0) =
e^{-\frac{M_0^2-M_1^2}{2\ M_0^2\ M_1^2} k^2} {\cal P}_{\it phy}({\vec
k}_1;M_0).
\end{equation}
If we keep the arbitrary scale ${\it l}_0$ fixed for comparison and let
the resolution scale ${\it l}_1 = M^{-1}_1 \gg {\it l}_0$ to run,
${\cal P}_{\it phy}({\vec k}_1;M_1) \simeq  e^{-\frac{k^2}{2\ M_1^2}}
{\cal P}_{\it phy}({\vec k}_1;M_0)$, we clearly see that the resolution
scale simply introduces a running physical ultraviolet cutoff $M_1$,
beyond which the power spectrum of the density fluctuations is
exponentially suppressed.

With these tools at hand we can obtain the variance
(\ref{variance}) of physical density anisotropies $(\Delta M_V[M])^2 =
\int \frac{d^3{\vec k}}{(2\pi)^3} {\cal P}_{\it phy}({\vec k}; M)
|F_V({\vec k})|^2$ as a function of the scale ${\it l} = M^{-1}$ at which
the density field (\ref{smooth}) is resolved. The important feature
to be noticed is that, for whatever two resolution scales $M_1$,$M_0$ that
we choose to compare, the normalization condition (\ref{normalization})
fixes ${\widetilde W}_{k = 0}[M_1,M_0] = 1$ at the origin $k = 0$ in
momentum space, see eq. (\ref{windowfunction}). Hence, the power spectrum 
at the origin is invariant ${\cal P}_{\it phy}(k = 0;M_1) = {\widetilde 
W}_{k = 0}[M_1,M_0]\ {\cal P}_{\it phy}(k = 0;M_0) = {\cal P}_{\it phy}(k 
= 0;M_0)$ under running of the resolution scale $M$. In particular,

\begin{equation}
\label{constraint}
{\cal P}_{\it phy}(k = 0;M_1) = 0 \hspace{0.3in} \Leftrightarrow
\hspace{0.3in} {\cal P}_{\it phy}(k = 0;M_0) = 0.
\end{equation}
According to eq. (\ref{equivalence1}), ${\cal P}(k = 0) = 0$ is a
necessary and sufficient condition for the variance of mass (energy)
anisotropies in a finite sub-volume $V$ to be scale invariant.
Let us remind that $d{\cal P}(k)/dk|_{k = 0} = 0$ is necessarily
fulfilled if ${\cal P}(k = 0) = 0$, up to some technical considerations.
Therefore, we must conclude that scale invariance (\ref{scaleinvariance})
of mass (energy) anisotropies in finite macroscopic sub-volumes
is an invariant feature of the statistical density anisotropies under
transformations of the scale ${\it l}$ at which we resolve the density
field (\ref{density}), as long as this scale is typically shorter than
the size $L$ of the considered sub-volume ${\it l} \simlt L$.

In eq. (\ref{implication}) we related the scale invariance of mass
(energy) anisotropies in macroscopic sub-volumes to statistical
fluctuations of the density field (\ref{smooth}) constrained to conserve
the total mass (energy), $(\Delta M_{\Omega})^2 = 0$. Equation
(\ref{constraint}) above simply expresses that this global constrain
should remain unchanged when we change the resolution scale at which we
test the density field.

It is interesting at this point to bring back our analysis of Section III,
case I. There we found a general approximate analytic expression for the
variance of scale invariant mass (energy) anisotropies in finite
macroscopic sub-volumes of size $L$, after introducing by hand a
regularization scale $k_c$:

\begin{equation}
\label{integral2H}
(\Delta M_V)^2 \sim 4 \rho^2_0\ L^3 \int_{0}^{k_c L}
d(k L) \frac{{\cal P}(k)}{(k L)^2} = 4 \rho^2_0\ L^2 \int_{0}^{k_c}
dk \frac{{\cal P}(k)}{k^2},
\end{equation}
which we wrote in (\ref{parameterRENORM}) as $(\Delta M_V)^2 = \rho^2_0\
\mu S$, with $\mu = 4 \int_{0}^{k_c} dk \frac{{\cal P}(k)}{k^2}$.
We found that the parameter $\mu$ is formally infinite when we take the
regularization scale $k_c$ to be infinitely large and noticed in Section
VI that this divergent parameter can be renormalized to its physical value
following an appropriate procedure. It is also worth to comment that in
our discussion in Section III the regularization scale $k_c$ was
introduced as a sharp ultraviolet cutoff in the integral above, but we
remarked there and in Fig. 1 that similar results are obtained when we
use, instead, exponential or polynomial cutoff procedures.

We now find exactly the same formal expression for the
renormalized (physical) variance of mass (energy) anisotropies in finite
sub-volumes $V$ of typical comoving size $L$ when the fluctuating density
field is probed to a length scale ${\it l} = M^{-1}$.

\begin{equation}
\label{integral42H}
(\Delta M_V)^2[M] \sim 4 \rho^2_0\ L^3 \int^{M L}
d(k L) \frac{{\cal P}_{\it phy}(k)}{(k L)^2} = 4 \rho^2_0\ L^2
\int_{0}^{M} dk \frac{{\cal P}_{\it phy}(k)}{k^2},
\end{equation}
In this expression the upper limit in the integral is not a sharp
ultraviolet cutoff, but it simply means that the resolution scale
introduces a physical exponential cutoff (\ref{PSR}) at $k \simgt M$.
This expression is obvioulsy finite. It describes a physically observable
magnitude (a dressed magnitude in the common vocabulary of
renormalization group equations). It is also clear from this expression
that the effect of changing the resolution at which we probe the mass
(energy) anisotropies can be reabsorbed in the renormalizable running
parameter $\mu_{\it phy}(M) = 4 \int_{0}^{M} dk \frac{{\cal P}_{\it
phy}(k)}{k^2}$, which measures the typical size of the scale invariant
anisotropies at that resolution scale \footnote{We are assuming at this
point that the considered sub-volume $V$ is bounded by a smooth enough
surface, such that its area $S(V)$ does not change with the resolution
scale at which we probe the 3D space.}, $(\Delta M_V)^2[M] \simeq
\rho^2_0\ \mu_{\it phy}(M) S(V)$, such that

\begin{equation}
\label{renormEQ}
\frac{d\mu_{\it phy}}{d{\it l}}|_{\it l} = -M^2\ \frac{d\mu_{\it 
phy}}{dM}|_M = -4\ {\cal P}_{\it phy}(M).
\end{equation}

This equation is a renormalization equation in the most common
sense, but it deserves an explanation. We have discussed in Section III
that all power spectra with index $n \ge 1$ are physically
indistinguishable through the scale invariant anisotropies
$(\Delta M_V)^2[M]$ they produce over finite, but macroscopically
large, volumes $V$ because, we said, the different indices $n$ can be
reabsorbed in the renormalized parameter $\mu_{\it phy}$.
The analysis in this section corroborates this conclussion and it offers
us a deeper insight. At a given resolution scale all the information
contained in any power spectra with index $n \ge 1$ is hidden in
the renormalized parameter $\mu_{\it phy}$, which measures the overall 
size of the scale invariant physical anisotropies. As we cannot predict
the physical value of this parameter in the theoretical setup that we
have laid, different power spectra with $n > 1$ are physically
indistingishable at this point.
On the other hand, the spectral index $n$ can be physically observable
through the renormalization equation (\ref{renormEQ}), that is, through
the change in the overall size of the scale invariant mass (energy)
anisotropies when we let the resolution scale $M$ to run.
We must conclude that, even though all power spectra with index $n \ge 1$
produce scale invariant mass (energy) anisotropies over macroscopically
large volumes, different spectral indices are physically distinguishable
if we can reliably measure the change $d\mu_{\it phy}/d{\it l}$ in the 
overall size of the anisotropies over a given macroscopic volume when we 
change the resolution scale ${\it l}$. This is the reason why in this 
section we have labelled the power spectrum ${\cal P}_{\it phy}(k)$ as a 
physically observable magnitude.

{\bf An additional note.} 
In the preceding paragraphs we have discussed how the resolution scale 
at which the random density field (\ref{smooth}) is probed affects 
the theoretical predictions on physically observable mass (energy) 
anisotropies. The analysis led us to the renormalization equation 
(\ref{renormEQ}). Quite naturally we found that the resolution 
scale introduces an exponential physical cutoff that suppresses all 
fluctuating modes with comoving wavelength shorter than the chosen scale 
of resolution. If this comoving scale is conveniently shorter than the 
comoving size of cosmological volumes the observed mass (energy) 
anisotropies are necessarily scale invariant. 

Before closing the discussion on this issue we want to comment on a 
different approach about how to manage the dependence of physical 
observables on the finite resolution scale of any real probe. This second 
approach is widely discussed in the literature on primordial anisotropies, 
but it is somehow artificial.

Let $\rho_{\it phy}({\vec x})$ be the physical density field resolved at
a certain scale ${\it l}_0$. The mass in a macroscopic sub-volume $V$ of
typical size $L \gg {\it l}_0$ (which we assume for simplicity centered
at the origin ${\vec x}=0$) is defined in this second approach as $M_{\it 
gauss}(V) = {\cal N} \int_{\Omega} d^3{\vec x} \rho_{\it phy}({\vec x}) 
e^{-|{\vec x}|^2/L^2}$, where ${\cal N}$ normalizes the gaussian weight, 
instead of the definition $M(V) = \int_V d^3{\vec x} \rho_{\it phy}({\vec 
x})$ we introduced in Section II and have used throghout the paper. The 
new operator $M_{\it gauss}(V)$ is commonly referred in the literature as 
the mass function with a normalized gaussian window, while $M(V)$ is 
commonly referred as a top-hat mass function. 

In Section VIII we have discussed how to properly accomodate a finite 
resolution scale within the standard definition of a top-hat mass 
function. The use of a mass function with a gaussian window 
$M_{\it gauss}(V)$ follows a different, and more artificial, approach: it 
averages the density field over the whole 3D space using a weighting 
gaussian function ${\cal N}\ e^{-x^2/L^2}$ with a typical opening width of 
the size $L$ of the considered volume $V$. 
According to eq. (\ref{smooth}) the gaussian mass function $M_{\it 
gauss}(V) \propto V \rho_{\it phy}({\vec x}=0;L)$ is nothing but the 
average mass in volumes of size $L$ when the physical density field is 
probed with a resolution scale of the same size of the volume $V$ of 
integration, ${\it l} \sim L$. 
The aim behind the definition $M_{\it gauss}(V)$ is to introduce in the
theoretical expression (\ref{integral42H}) for the variance of mass
anisotropies in a sub-volume of typical size $L$ an ultraviolet cutoff
in momentum space of the order $k \sim 1/L$ and artificially justify
the estimation (\ref{berlusco}). It is important to notice that 
within this gaussian approach the volume $V$ is not treated as a 
macroscopic volume, but as a volume of the minimal size that the probe 
can test. And we must be aware that a theory endowed with an ultraviolet 
cutoff in real space of size ${\it l}$ can lead to contrieved
predictions when applied to physical sub-systems whose typical size is $L 
\simlt {\it l}$, in the same sense that the effective Fermi four-point 
vertex fails to reproduce the observational features of weak interactions 
at energies close or higher than the mass of the $W$ and $Z$ bosons.
In both cases, we might be trying to apply the theory beyond the scales it 
can reach. In the case of a power spectrum with index $n = 1$ the result 
of using the gaussian mass function $M_{\it gauss}(V)$ does not modify the 
scale invariance (\ref{renormalization}), but it artificially render the
divergent parameter $\mu$ in (\ref{parameterRENORM}) finite. In the case 
of a power spectrum with index $n > 1$ the use of a gaussian mass function 
artificially produce the wrong results (\ref{preliminaries}), instead of 
the scale invariant (\ref{renormalization}).

\section{Discussion.}

We have revised in this paper the standard theoretical framework 
commonly used to discuss the origin of scale invariant cosmological mass 
(energy) anisotropies in the homogeneous and isotropic FRW background at 
the time of matter radiation equality to point out a crucial fault 
(\ref{preliminaries}) in the argument that led to wrongly assert the 
so-called origin of structures problem of standard cosmology. We have 
shown that the correct evaluation (\ref{tres}) offers a natural and 
elegant explanation of the scale invariance of the primordial 
cosmological anisotropies as a consequence of a conservation 
law (\ref{implication}) of the total mass (energy) of the fluctuating 
density field. Surprisingly, the correct estimation (\ref{tres}) that we 
present here is well-known since long ago in statistical mechanics and 
condensed matter physics, but was never properly noticed before in cosmology.

According to our analysis primordial cosmological mass (energy) 
anisotropies in the, otherwise, homogeneous, isotropic and flat FRW 
universe at the time of matter radiation equality could happen to be 
simply quantum vacuum fluctuations of the density field generated at 
the same instant of equality with comoving wavelength shorter than the 
causal horizon at that time. In other words, the mass (energy) 
primordial anisotropies over cosmologically large volumes would happen 
randomly as a result of local rearrangement of matter at the instant of 
equality through the surface that bounds the considered spatial regions, 
thus explaining why their variance is proportional to the area of the 
boundary surface.

Obviously, this explanation of the origin of primordial cosmological 
anisotropies at the time of matter radiation equality in the context of 
standard cosmology does not require any previous epoch of inflationary 
expansion. In the context of this alternative scenario the fluctuating 
density field is naturally required to be in its ground state, thus 
avoiding also the problematic issues on the exceptional initial 
conditions of the universe that can produce a period of inflationary 
expansion \cite{Hollands:2002yb}. Furthermore, the physics involved in 
the generation of the anisotropies shall probably be physics at the scale 
$T_{eq} \sim 1$~eV of equality, instead of the very high energy scales 
$T \simgt 10^{15}$~GeV summoned in inflationary mechanism.  
Yet, this alternative mechanism to the origin of large scale cosmological 
structures needs a further elaboration of the calculations presented in 
the last section VII on the interplay of the surface mass (energy) 
anisotropies with the theory of gravitational structure formation beyond 
the linear approximation.

Finally, we must remark that the correct estimation (\ref{tres}), which 
lies at the foundation of all the arguments in this paper, implies a 
continuous degeneracy in the power spectrum (\ref{variance}) of 
statistical systems that produce scale invariant mass (energy) 
anisotropies, which we expressed in our discussion through the 
impossibility to distinguish spectra with indexes $n > 1$ over 
cosmologically large comoving volumes. The degeneracy is associated to 
the possibility of transforming the charge density field operator of a 
globally conserved Noether charge without actually modifying the charge
and it has far reaching consequences.
Such transformations do not modify the scale invariance of the 
spatial quantum anisotropies in the distribution of the charge, but they 
can actually change the formal expression of the infinite parameter that 
measures the size of these anisotropies. Or, from a different perspective
these transformations can be reabsorved in the definition of an infinite
renormalizable parameter of the theory. From this point of view we 
understand the scale invariance statement (\ref{renormalization})
in the sense of a renormalization group equation and conclude that
the physical dressed parameter, {\it i.e.} the absolute size of the
anisotropies, cannot be predicted in the available setup of quantum field 
theories. In the same sense that we cannot predict the actual value of 
coupling constants or masses, or the cosmological constant.

The picture of the origin of primordial cosmological fluctuations as 
vacuum fluctuations at the same instant of matter radiation equality would
not be complete without a discussion of how vacuum fluctuations can
decohere. Halliwell, in \cite{Halliwell:1998jf}, advanced that in the
context of the decoherent histories approach to quantum mechanics
\cite{Gell-Mann:kh} fluctuations of local densities
(momentum, energy) are more prone to decohere. More recently we have
developed a new formalism \cite{Brustein:2002zn} to address
the same issue: in this formalism the ground state of a bosonic or
fermionic system is described as a linear combination of randomly
distributed pseudoclassical incoherent paths, which allows a description 
of the fluctuations of {\it collective} operators like the energy in a
sub-volume in terms of classical stochastic concepts.

\section{Acknowledgments.}

I am glad to thank J.~Oaknin for his suggestions.
I am thankful to A.~Yarom and R.~Brustein for discussions on a 
related project. I am also thankful to M.~Joyce and D.~Chung for their 
comments on an earlier version of this paper.
This work was partly supported by the National Science and Engineering 
Research Council of Canada. I am grateful to Ana Oaknin, my 
teacher, tutor and friend, in memoriam.

\end{document}